\documentclass{ws-ijmpd}
\usepackage[super,compress]{cite}
\usepackage{graphicx}
\renewcommand{\thefigure}{\thesubsection.\arabic{figure}}
\begin{document}
\markboth{Gogberashvili, Mantidze, Sakhelashvili and Shengelia}{Standing Wave Braneworlds}

\catchline{}{}{}{}{}
%
%%%%%%%%%%%%%%%%%%%%%%%%%%%%%%%%%%%%%%%%%%%%%%%%%%%%%%%%%%%%%%%%%%%%

\title{Standing Waves Braneworlds}

\author{\bf Merab Gogberashvili}
\address{Javakhishvili Tbilisi State University, 3 Chavchavadze Ave\\
Tbilisi 0179, Georgia;\\
Andronikashvili Institute of Physics, 6 Tamarashvili St.,\\
Tbilisi 0177, Georgia\\
gogber@gmail.com}

\author{\bf Irakli Mantidze}
\address{Javakhishvili Tbilisi State University, 3 Chavchavadze Ave\\
Tbilisi 0179, Georgia\\
Irakli.Mantidze@ens.tsu.edu.ge}

\author{\bf Otari Sakhelashvili}
\address{Javakhishvili Tbilisi State University, 3 Chavchavadze Ave\\
Tbilisi 0179, Georgia\\
otosaxel@gmail.com}

\author{\bf Tsotne Shengelia}
\address{Javakhishvili Tbilisi State University, 3 Chavchavadze Ave\\
Tbilisi 0179, Georgia\\
CotneShengelia@yahoo.com}

\maketitle

\begin{history}
\received{Day Month Year}
\revised{Day Month Year}
\end{history}

\begin{abstract}
The class of non-stationary braneworld models generated by the coupled gravitational and scalar fields is reviewed. The model represents a brane in a space-time with single time and one large (infinite) and several small (compact) space-like extra dimensions. In some particular cases the model has the solutions corresponding to the bulk gravi-scalar standing waves bounded by the brane. Pure gravitational localization mechanism of matter particles on the node of standing waves, where the brane is placed, is discussed. Cosmological applications of the model also considered.

\end{abstract}

\ccode{PACS numbers: 04.62.+v; 11.27.+d; 98.80.Cq}

%%%%%%%%%%%%%%%%%%%%%%%%%%%%%%%%%%%%%%%%%%%%%%%%%%%%%%%%%%%%%%%%%%%%%%%%%

\tableofcontents

%%%%%%%%%%%%%%%%%%%%%%%%%%%%%%%%%%%%%%%%%%%%%%%%%%%%%%%%%%%%%%%%%%%%%%%%%

\section{\bf Introduction}

The models with large extra dimensions involving 3-dimensional singular space-like surfaces with non-factorizable geometry, braneworlds \cite{brane-1, brane-2, brane-3, brane-4}, have attracted a lot of interest recently (see \cite{rev-1, rev-2, rev-3, rev-4} for reviews). A key requirement for realizing the braneworld idea is that the various matter fields be localized on the brane. It is preferable to have a universal gravitational trapping mechanism for all fields. However, there are difficulties to realize such mechanism with the exponentially warp factor used in standard brane scenarios. In the existing (1+4)-dimensional models spin $0$ and spin $2$ fields are localized on the brane with the decreasing warp factor \cite{brane-1, brane-2, brane-3, brane-4}, spin $1/2$ field can be localized with the increasing warp factor \cite{BaGa}, and spin $1$ fields are not localized at all \cite{Po}. For the case of (1+5)-dimensions it was found that spin $0$, spin $1$ and spin $2$ fields are localized on the brane with the decreasing warp factor and spin $1/2$ fields again are localized with the increasing warp factor \cite{Od}. There exist also 6D models with non-exponential warp factors that provide gravitational localization of all kind of bulk fields on the brane \cite{6D-old-1, 6D-old-2, 6D-old-3, 6D-old-4, 6D-old-5}, however, these models require introduction of unnatural sources.

To solve the localization and some other problems of the braneworlds with static geometric configurations there have appeared models which use time-dependent metrics \cite{S-1, S-2, S-3, S-4}. One such approach is proposed recently the standing wave braneworld model with gravi-scalar waves in the bulk \cite{5D-ghost-1, 5D-ghost}. This kind of models can provide a natural alternative mechanism for universal gravitational trapping of zero modes of all kinds of matter fields.

To clarify the mechanism of localization used in standing waves braneworlds let us remind that standing electromagnetic waves, so-called optical lattices, can provide trapping of various particles by scattering, dipole and quadruple forces \cite{Opt-1, Opt-2, Opt-3, Opt-4}. It is known that the motion of test particles in the field of a gravitational wave is similar to the motion of charged particles in the field of an electromagnetic wave \cite{Ba-Gr}. Thus standing gravitational waves could also lead to confinement of matter (via quadruple forces). For example, the equations of motion of a system of spinless particles in the quadruple approximation has the form \cite{Dix}:
\begin{equation} \label{quad}
\frac{Dp^\mu}{ds}= F^\mu = -\frac 16 J^{\alpha\beta\gamma\delta}D^\mu R_{\alpha\beta\gamma\delta}~,
\end{equation}
where $p^\mu$ is the momentum and $J^{\alpha\beta\gamma\delta}$ is the quadruple moment of the stress-energy tensor of the matter. The oscillating metric due to gravitational waves should induce a quadruple moment in the matter fields. If the induced quadruple moment is out of phase with the gravitational wave the system energy increases and the particles will feel a quadruple force, $F^\mu$, which ejects them out of the high curvature region towards the nodes of standing waves.

In this paper we review existing standing waves braneworld models in various dimensions. The paper is organized as follows. Sec. \ref{General} presents the main metric {\it ansatz} and general $N$-dimensional setup of the model. In Sec. \ref{k=0}, \ref{S=0}, \ref{k ne 0} and \ref{6D normal} some exact solutions to the system of Einstein and Klein-Gordon equations in various dimensions are considered. Sec. \ref{Cosmology} discusses cosmological applications of standing waves braneworlds. In \ref{Averages} the formulas for time averaged oscillatory functions, used throughout the paper, are presented.

%%%%%%%%%%%%%%%%%%%%%%%%%%%%%%%%%%%%%%%%%%%%%%%%%%%%%%%%%%%%%%%%%%%%%%%%%

\section{\bf General setup}
\label{General}

Standing waves braneworlds are realized as wave solutions to the system of Einstein and Klein-Gordon equations. The setup consists of a single brane and non-self-interacting scalar field, $\phi$, in multi-dimensional space with single time-like dimension and the signature $(+,-, -,. . . ,-)$. The action of the model in $N$-dimensional case has the form \cite{GMT}:
\begin{equation}\label{action-N}
S = \int d^Nx \sqrt {|g|} \left( \frac{M^{N-2}}{2}R + \Lambda + \frac \epsilon 2 g^{AB}\partial _A \phi \partial _B\phi  + L_{brane} \right)~,
\end{equation}
where $\Lambda$ denotes the bulk cosmological constant, $L_{brane}$ is the brane Lagrangian and $M$ is the fundamental scale, which is related to the $N$-dimensional Newton constant, $G=1/(8\pi M^{N-2})$. The sign coefficient $\epsilon$ in front of the Lagrangian of $\phi$ takes the values $+1$ and $-1$ for the real and phantom bulk scalar fields, respectively. Capital Latin indexes numerate $N$-dimensional coordinates, and we use the units where $c = \hbar = 1$.

Variation of the action (\ref{action-N}) with respect to $g_{AB}$ leads to the Einstein equations:
\begin{equation}\label{EinsteinEq}
R_{AB} - \frac 12 g_{AB}R = \frac {1}{M^{N-2}}(\sigma_{AB} + \epsilon T_{AB})~,
\end{equation}
where the source terms are the energy-momentum tensors of the bulk scalar field,
\begin{equation}\label{ScalarFieldEnergyMomentumTensor}
T_{AB} = \partial _A\phi \partial _B\phi - \frac 12 g_{AB} \partial ^C\phi \partial _C\phi~,
\end{equation}
and of the brane,
\begin{equation} \label{sigma}
\sigma _B^A = M^{N-2}\delta (z)\mathrm{diag} \left[\tau_t,\tau_{x_1},...,\tau _{x_{(N - 3)}},\tau_y,\tau_z\right]~,
\end{equation}
with $\tau_A$ being brane tensions. For the sources (\ref{ScalarFieldEnergyMomentumTensor}) and (\ref{sigma}) the Einstein equations (\ref{EinsteinEq}) can be rewritten in the form:
\begin{equation}\label{EinsteinEquations4}
R_{AB}= \frac {1}{M^{N-2}}\left(\sigma_{AB}-\frac{1}{N-2}g_{AB}\sigma + \epsilon \partial_A \phi \partial_B \phi \right)~.
\end{equation}

The solution to (\ref{EinsteinEquations4}), which generates standing wave braneworlds, has the form \cite{GMT}:
\begin{equation}\label{MetricAnsatzGeneral}
ds^2 = (1 + k|z|)^c e^S \left(dt^2 - dz^2\right) - (1 + k|z|)^b \left[e^V \sum\limits_{i = 1}^{N - 3} dx_i^2 + e^{B - (N - 3)V}dy^2 \right]~,
\end{equation}
where $c$, $b$ and $k$ are some constants, and the metric functions $S = S(t,|z|)$, $V = V(t,|z|)$ and $B = B(t,|z|)$ depend only on time, $t$, and on the modulus of the orthogonal to the brane extra coordinate $z$.

The metric (\ref{MetricAnsatzGeneral}) describes geometry of the $(N-1)$-brane placed at the origin of the large space-like extra dimension $z$. Among the $(N-2)$ remaining spatial coordinates, three: $x_1$, $x_2$ and $y$, denote the ordinary infinite dimensions of our world, while $x_i$ ($i = 3, ..., N-5$) is assumed to be compact, curled up to the unobservable sizes for the present energies. Note that the compact dimensions also are brane coordinates for $z=0$. This particular feature is called hybrid compactification \cite{hybrid}.

Most of the standing wave braneworld models assume
\begin{equation}
B(t,|z|) = 0
\end{equation}
in (\ref{MetricAnsatzGeneral}), since in this case braneworld solutions can be found for symmetric bulk cosmological constant. For simplification of classification we shall mostly use the metric (\ref{MetricAnsatzGeneral}) without the factor $e^B$. The only standing wave braneworld with $B(t,|z|) \ne 0$ considered in the literature will be reviewed in Sec. \ref{6D normal}.

Using the expression of the determinant for (\ref{MetricAnsatzGeneral}) with $B(t,|z|) = 0$,
\begin{equation}\label{DeterminantOfMetric}
\sqrt {|g|} = e^S (1 + k|z|)^{\frac{b(N - 2) + 2c}{2}}~,
\end{equation}
the Klein-Gordon equation for the bulk scalar field, $\phi$, in the background metric (\ref{MetricAnsatzGeneral}) takes the form:
\begin{eqnarray}\label{BulkScalarFieldEquation}
&&\left[ \partial _t^2 - \partial_z^2 + \frac{b(N - 2) k ~\mathrm{sgn} (z)}{2( 1 + k|z|)} \partial_z - \right.\nonumber \\
&-& \left. (1 + k|z|)^{c - b} e^S \left( e^{ - V}\sum \limits_{i = 1}^{N - 3} \partial _{x_i}^2 + e^{(N - 3)V}\partial _y^2 \right)\right]\phi = 0~,
\end{eqnarray}
where $\mathrm{sgn} (z)$ is the sign function.

The non-zero components of $N$-dimensional Ricci tensor for the metric {\it ansatz} (\ref{MetricAnsatzGeneral}) with $B(t,|z|) = 0$ are:
\begin{eqnarray}\label{RicciTensorComponents1}
R_{tt} &=& \left(ck + S'\right)\delta (z) + \frac 12\left[ S'' + \frac{b(N - 2)k}{2(1 + k|z|)}S' - \ddot S - \frac{(N - 2)(N - 3)}{2}\dot V^2 \right] + \nonumber \\
&+& \frac{c[b(N - 2) - 2]k^2}{4(1 + k|z|)^2}~,\nonumber\\
R_{tz} &=& \frac{b(N - 2)k~\mathrm{sgn} (z)}{4(1 + k|z|)}\dot S - \frac{(N - 2)(N - 3)~\mathrm{sgn}(z)}{4}\dot VV'~, \nonumber\\
R_{x_1x_1} &=& ... = R_{x_{(N - 3)}x_{(N - 3)}} =  - \left(bk + V'\right)e^{- S + V}\delta (z) +  \\
&+& \frac{e^{- S + V}}{(1 + k|z|)^{c - b}}\left\{ \frac 12\left[\ddot V - V'' - \frac{b(N - 2)k}{2(1 + k|z|)}V' \right] - \frac{b[b(N - 2) - 2]k^2}{4(1 + k|z|)^2} \right\}~ , \nonumber\\
R_{yy} &=& - \left[bk - (N - 3)V'\right]e^{- S - (N - 3)V}\delta (z) - \nonumber\\
&-& \frac{e^{- S - (N - 3)V}}{(1 + k|z|)^{c - b}}\left\{\frac{(N - 3)}{2}\left[\ddot V - V'' - \frac{b(N - 2)k}{2(1 + k|z|)}V' \right] + \frac{b[b(N - 2) - 2]k^2}{4(1 + k|z|)^2} \right\}~, \nonumber \\
R_{zz} &=& - \left\{ \left[ c + b(N - 2)\right]k + S' \right \} \delta (z) +  \nonumber \\
&+& \frac 12\left[ \ddot S - S'' + \frac{b(N - 2)k}{2(1 + k|z|)}S' - \frac{(N - 2)(N - 3)}{2}V'^2 \right] + \frac{Dk^2}{4(1 + k|z|)^2} ~, \nonumber
\end{eqnarray}
where overdots and primes denote the derivatives with respect to $t$ and $|z|$, respectively, and to shorten the last expression we have introduced the constant:
\begin{equation}
D = c[b(N - 2) + 2] - b(N - 2)(b - 2)~.
\end{equation}

The Einstein equations (\ref{EinsteinEquations4}) can be split into the system of equations for metric functions:
\begin{eqnarray}\label{SystemOfEquationsForMetricFunction}
\frac 12\left[ S'' + \frac{b(N - 2)k}{2(1 + k|z|)}S' - \ddot S - \frac{(N - 2)(N - 3)}{2}\dot V^2 \right] + \frac{c[b(N - 2) - 2]k^2}{4(1 + k|z|)^2} = \nonumber \\
= \epsilon \frac{1}{M^{N-2}}\partial _t\phi^2~, \nonumber\\
\frac{b(N - 2)k~\mathrm {sgn}(z)}{4(1 + k|z|)}\dot S - \frac{(N - 2)(N - 3)~\mathrm {sgn}(z)}{4}\dot VV' = \nonumber \\
= \epsilon \frac{\mathrm {sgn}(z)}{M^{N-2}}\partial _t\phi \partial _z\phi~, \nonumber \\
\frac{e^{- S + V}}{(1 + k|z|)^{c - b}}\left\{\frac 12\left[ \ddot V - V'' - \frac{b(N - 2)k}{2(1 + k|z|)V'} \right] - \frac{b[b(N - 2) - 2]k^2}{4(1 + k|z|)^2} \right\} = \nonumber \\ = \epsilon \frac{1}{M^{N-2}}\partial_{x_1}\phi^2~, \nonumber \\
... ~~~~~~~~~~ \\
\frac{e^{- S + V}}{(1 + k|z|)^{c - b}}\left\{ \frac 12 \left[\ddot V - V'' - \frac{b(N - 2)k}{2(1 + k|z|)}V' \right] - \frac{b[b(N - 2) - 2]k^2}{4(1 + k|z|)^2} \right\} = \nonumber \\ = \epsilon \frac{1}{M^{N-2}}\partial_{x_{(N - 3)}}\phi^2~, \nonumber \\
\frac{e^{- S - (N - 3)V}}{(1 + k|z|)^{c - b}}\left\{ \frac{(N - 3)}{2}\left[V'' - \ddot V + \frac{b(N - 2)k}{2(1 + k|z|)}V' \right] - \frac{b[b(N - 2) - 2]k^2}{4(1 + k|z|)^2} \right\} = \nonumber \\
= \epsilon \frac{1}{M^{N-2}}\partial_y\phi^2~, \nonumber \\
\frac 12 \left[\ddot S - S'' + \frac{b(N - 2)k}{2(1 + k|z|)}S' - \frac{(N - 2)(N - 3)}{2}V'^2 \right] + \frac{Dk^2}{4(1 + k|z|)^2} = \nonumber \\
=\epsilon \frac{1}{M^{N-2}}\partial_z\phi^2~, \nonumber
\end{eqnarray}
and for the brane energy-momentum tensor:
\begin{eqnarray}\label{SystemOfEquationsForBraneEnergyMomentumTensor}
\left( ck + S'\right)\delta (z) = \frac{1}{M^{N-2}}\left( \sigma _{tt} - \frac{1}{N - 2}~g_{tt}\sigma \right)~, \nonumber\\
- \left( bk + V'\right)e^{- S + V}\delta (z) = \frac{1}{M^{N-2}}\left(\sigma _{x_1x_1} - \frac{1}{N - 2}~g_{x_1x_1}\sigma  \right)~, \nonumber\\
  ... ~~~~~~~~~~\\
- \left( bk + V' \right)e^{- S + V}\delta (z) = \frac{1}{M^{N-2}}\left(\sigma _{x_{(N - 3)}x_{(N - 3)}} - \frac{1}{N - 2}~g_{x_{(N - 3)}x_{(N - 3)}}\sigma \right)~, \nonumber\\
- \left[bk - (N - 3)V'\right]e^{- S - (N - 3)V}\delta (z) = \frac{1}{M^{N-2}}\left( \sigma _{yy} - \frac{1}{N - 2}~g_{yy}\sigma \right)~, \nonumber\\
- \left\{ [c + b(N - 2)]k + S' \right\} \delta (z) = \frac{1}{M^{N-2}}\left( \sigma _{zz} - \frac{1}{N - 2}~g_{zz}\sigma  \right)~. \nonumber
\end{eqnarray}

In the following sections \ref{k=0}, \ref{S=0}, \ref{k ne 0} and \ref{6D normal} we present different solutions to the system (\ref{BulkScalarFieldEquation}), (\ref{SystemOfEquationsForMetricFunction}) and (\ref{SystemOfEquationsForBraneEnergyMomentumTensor}) in various dimensions.

%%%%%%%%%%%%%%%%%%%%%%%%%%%%%%%%%%%%%%%%%%%%%%%%%%%%%%%%%%%%%%%%%%%%%%%%%

\section{\bf Solutions with $k=0$}
\label{k=0}

We start with the case $k=B=0$ when the general metric {\it ansatz} (\ref{MetricAnsatzGeneral}) obtains the form:
\begin{equation}\label{MetricAnsatzGeneral2}
ds^2 = e^S \left(dt^2 - dz^2\right) - e^V \sum\limits_{i = 1}^{N - 3} dx_i^2 - e^{- (N - 3)V}dy^2 ~.
\end{equation}
The metric functions $S(t,|z|)$ and $V (t,|z|)$ are depending on the modulus of the extra dimension coordinate $z$. The Ricci tensor has the $\delta$-like singularity at $z=0$ and to smooth it the brane is placed at the origin of $z$. Note that without modulus for $z$ the metric (\ref{MetricAnsatzGeneral2}) will correspond to the running wave solutions, considered in \cite{Yurt, Fe-Ib, Griff} for 4D case.

%%%%%%%%%%%%%%%%%%%%%%%%%%%%%%%%%%%%%%%%%%%%%%%%%%%%%%%%%%%%%%%%%%%%%%%%%

\subsection{The oscillating brane}

For the simplest case,
\begin{equation}
k = \phi = V = 0~,
\end{equation}
the solution to the system (\ref{BulkScalarFieldEquation}) and (\ref{SystemOfEquationsForMetricFunction}) is \cite{GMT}:
\begin{equation}
S = \left[ C_1\sin (\Omega t) + C_2\cos (\Omega t)\right]\left[ C_3\sin (\Omega |z|) + C_4\cos (\Omega |z|) \right]~,
\end{equation}
where $C_i$ ($i=1, 2, 3,4$) and $\Omega$ are the constants. This solution corresponds to the oscillating brane at $|z|=0$ in $N$-dimensional space-time. Imposing on this only nontrivial function the boundary condition,
\begin{equation}
S|_{|z| = 0} = 0 ~,
\end{equation}
from the equations (\ref{SystemOfEquationsForBraneEnergyMomentumTensor}) one can find also the brane tensions:
\begin{eqnarray}
\tau_t &=& \tau_z = 0~, \nonumber \\
\tau _{x_1} &=& \tau_{x_2} = ... = \tau_{x_{(N - 3)}} = - S'~, \\
\tau_y &=& - S'~.\nonumber
\end{eqnarray}

%%%%%%%%%%%%%%%%%%%%%%%%%%%%%%%%%%%%%%%%%%%%%%%%%%%%%%%%%%%%%%%%%%%%%%%%%

\subsection{4D gravi-scalar breather}

For the 4D variant of the metric (\ref{MetricAnsatzGeneral2}), in the case with normal scalar field, $\epsilon = +1$ in (\ref{action-N}), the system (\ref{BulkScalarFieldEquation}) and (\ref{SystemOfEquationsForMetricFunction}) has the standing waves solution \cite{GK},
\begin{eqnarray} \label{sol-4}
\phi (t,|z|) &=& f(|z|) \cos (\omega t)~, \nonumber \\
V (t,|z|) &=& f(|z|) \sin (\omega t), \\
S (t,|z|) &=& 2\int_0^\infty dz~ z\left[\omega^2 f^2(z) + f'^2(z)\right] - V(t,|z|)~, \nonumber
\end{eqnarray}
where $\omega$ is the oscillation frequency of waves and the function $f(|z|)$ is expressed by a zero order Bessel function of the first kind,
\begin{equation}
f(z) \sim J_0(\omega |z|)~.
\end{equation}

The coherent process described by (\ref{sol-4}) mimics very much the behavior of the electromagnetic waves in two-level media \cite{LAB1, LAB2}, where the electromagnetic waves are periodically absorbed and radiated by the two-level atoms, and thus are trapped all the time inside the medium. The solutions (\ref{sol-4}) describe gravitational waves bounded by the domain wall transferring periodically the energy to the matter (bulk scalar waves) and back.

%%%%%%%%%%%%%%%%%%%%%%%%%%%%%%%%%%%%%%%%%%%%%%%%%%%%%%%%%%%%%%%%%%%%%%%%%

\subsection{Gravi-ghost standing waves in $N$ dimensions}

For the case of the phantom bulk scalar field ($\epsilon = -1$), when the both metric functions $S$ and $V$ are presented in (\ref{MetricAnsatzGeneral2}), the system (\ref{BulkScalarFieldEquation}) and (\ref{SystemOfEquationsForMetricFunction}) has the standing wave solution of the form \cite{GMT}:
\begin{eqnarray}\label{Solution6}
V &=& \left[ C_1\sin (\omega t) + C_2\cos (\omega t) \right]\left[ C_3\sin (\omega |z|) + C_4\cos (\omega |z|) \right]~, \nonumber \\
\phi &=& \frac 12 \sqrt {M^{N-2}(N - 2)(N - 3)} \left[ C_1\sin (\omega t) + C_2\cos (\omega t) \right]\times \nonumber \\
&\times& \left[ C_3\sin ( \omega |z|) + C_4\cos (\omega |z|) \right], \\
S &=& \left[ C_5\sin (\Omega t) + C_6\cos (\Omega t) \right]\left[ C_7\sin ( \Omega |z|) + C_8\cos (\Omega |z|) \right]~,\nonumber
\end{eqnarray}
with $C_i$ ($i=1, 2, 3,...,8$), $\Omega$ and $\omega$ being some constants.

Imposing on the metric functions $S$ and $V$ the boundary conditions on the brane:
\begin{equation}
S|_{|z| = 0} = V|_{|z| = 0} = 0~,
\end{equation}
from (\ref{SystemOfEquationsForBraneEnergyMomentumTensor}) one can find the brane tensions:
\begin{eqnarray}
\tau_t &=& \tau _z = 0~, \nonumber\\
\tau_{x_1} &=& \tau_{x_2} = ... = \tau_{x_{(N - 3)}} = - S' + V'~,\\
\tau_y &=& - S' - (N - 3)V'~.\nonumber
\end{eqnarray}
It is clear from (\ref{Solution6}) that there are two different frequencies associated with the metric functions $S$ and $V$ ($\Omega$ and $\omega$, respectively), and that the oscillation frequency of the phantom bulk scalar field standing wave, unlike to the case of gravi-scalar breather considered in the previous paragraph, coincides with the frequency of the standing gravitational wave.

%%%%%%%%%%%%%%%%%%%%%%%%%%%%%%%%%%%%%%%%%%%%%%%%%%%%%%%%%%%%%%%%%%%%%%%%%

\subsection{6D standing wave braneworld with ghost scalars}

Consider the 6D version of the metric (\ref{MetricAnsatzGeneral2}) \cite{6D-ghost-oto}:
\begin{equation} \label{ansatz-6D}
ds^2 = e^S dt^2 - e^V \left(dx^2 + dy^2 + dz^2\right) - dr^2 - e^{-3V} d\theta^2~,
\end{equation}
where $x$, $y$ and $z$ denote coordinates of the 3-space along the brane, the large extra dimension is labeled by $r$, and the sixth coordinate $\theta$ is assumed to be compact. Advantage of this model is the isotropy of the 3-space of the brane under the oscillations $V(t,|r|)$, what is important in cosmological applications (see (\ref{ansatz})).

For the {\it ansatz} (\ref{ansatz-6D}) the system (\ref{BulkScalarFieldEquation}) and (\ref{SystemOfEquationsForMetricFunction}) has the standing wave solutions:
\begin{eqnarray}\label{Sol-6D}
S(|r|) &=& \ln \left(1 + \frac {|r|}{a} \right)^2~, \nonumber \\
V(t,|r|) &=& C \sin (\omega t) \sin \left( a\omega \ln\left[1+\frac{|r|}{a}\right]\right)~,
\end{eqnarray}
where $a$, $C$ and $\omega$ are the integration constants.

%%%%%%%%%%%%%%%%%%%%%%%%%%%%%%%%%%%%%%%%%%%%%%%%%%%%%%%%%%%%%%%%%%%%%%%%%

\subsubsection{Localization problem}

Consider the localization problem in the 6D space-time (\ref{ansatz-6D}) for the massless scalar field, $\Phi$, defined by the action \cite{6D-ghost-oto}:
\begin{equation}\label{Action-Phi}
S = \frac 12 \int d^{6}x \sqrt{g}g^{MN} \partial_M\Phi\partial_N\Phi~.
\end{equation}
When the frequency of bulk standing waves is much larger than frequencies associated with the energies of matter particles on the brane it is possible to perform time averaging of the oscillating exponents $e^V$ (see \ref{Averages}) and to separate the variables:
\begin{equation}\label{fsi}
\Phi(x^A) = \psi(t,x,y,z)\sum_l\nu_l(r)e^{il\theta}~.
\end{equation}
Consider the $S$-wave solution ($l=0$), i.e. assume that nothing depends on the extra dimension angle $\theta$. Then the time averaged action (\ref{Action-Phi}) takes the form:
\begin{eqnarray}\label{Action_6D_last}
S &=& \frac 12 \int d^6x\left[\frac{\nu^2 \partial_t\psi^2}{1 + r/a} - \left(1 + \frac{r}{a}\right)\nu'^{2}\psi^2 - \right. \nonumber \\
&-& \left. \left(1 + \frac{r}{a}\right)\nu^2 \left\langle e^V \right\rangle \left(\partial_x\psi^2 + \partial_y\psi^2 + \partial_z\psi^2 \right) \right]~.
\end{eqnarray}

In general, to have a field localized on a brane 'coupling' constants appearing after integration of an action over extra coordinates must be non-vanishing and finite. So normalizable zero modes of the scalar field $\Phi$ will exist on the brane if the action (\ref{Action_6D_last}) is integrable over $r$, i.e. the functions $\nu'^2(1 + r/a)$, $\left\langle e^V\right\rangle \nu^2 (1 + r/a)$ and $\nu^2/(1+r/a)$ are integrable.
%%%%%%%%%%%%%%%%%%%%%%%%%%%%%%%%%%%%%%%%%%%%%
\begin{figure}[ht]
\begin{center}
\includegraphics[width=0.7\textwidth]{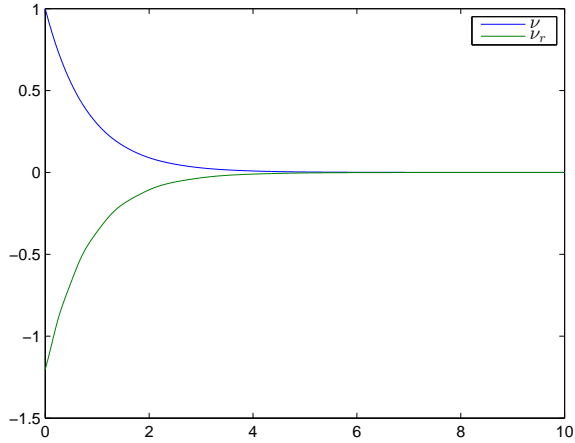}
\caption{Numerical solutions for $\nu$ and $\nu'$.}
\label{fig-sol}
\end{center}
\end{figure}
%%%%%%%%%%%%%%%%%%%%%%%%%%%%%%%%%%%%%%%%%%%

Fig. \ref{fig-sol} displays the solution to the 6D Klein-Gordon equation for extra dimension part of $\Phi$ and its first derivative close to the brane \cite{6D-ghost-oto}.

\begin{figure}[ht]
\begin{center}
\includegraphics[width=0.7\textwidth]{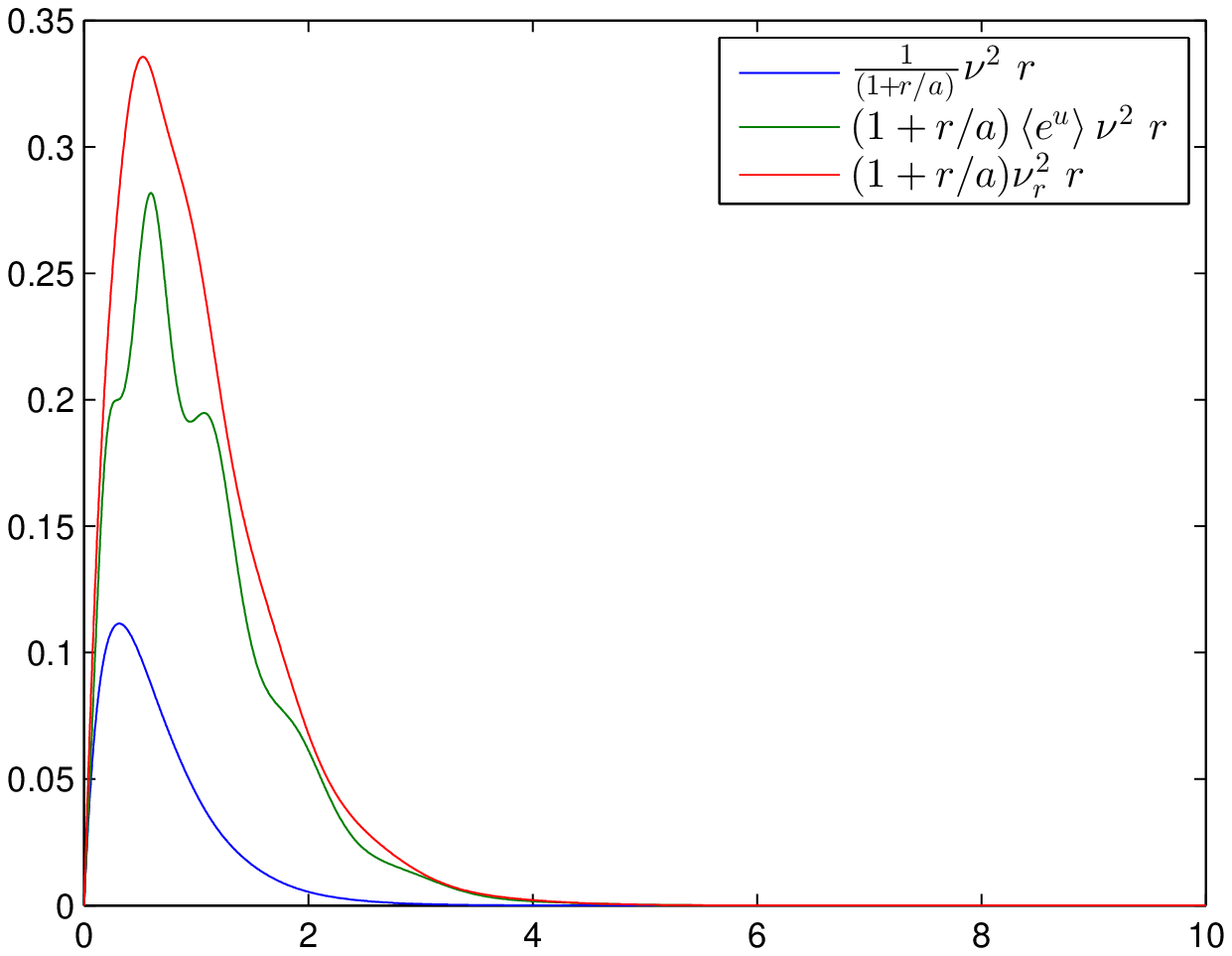}
\caption{$r$-depended factors in (\ref{Action_6D_last}) multiplied by $r$}
\label{fig-int}
\end{center}
\end{figure}

Fig. \ref{fig-int} shows that all $r$-depended factors in (\ref{Action_6D_last}) that are multiplied by the extra coordinate $r$ are decreasing functions, i.e. (\ref{Action_6D_last}) is integrable over $r$ and the scalar field zero modes are localized on the brane.

%%%%%%%%%%%%%%%%%%%%%%%%%%%%%%%%%%%%%%%%%%%%%
%%%%%%%%%%%%%%%%%%%%%%%%%%%%%%%%%%%%%%%%%%%%

%%%%%%%%%%%%%%%%%%%%%%%%%%%%%%%%%%%%%%%%%%%%%%%%%%%%%%%%%%%%%%%%%%%%%%%%%

\section{\bf Solutions with $b=c=2$ and $S = 0$}
\label{S=0}

In this section we consider the solutions to the system (\ref{BulkScalarFieldEquation}) and (\ref{SystemOfEquationsForMetricFunction}) when the main metric {\it ansatz} (\ref{MetricAnsatzGeneral}) can be written as:
\begin{equation}\label{MetricAnsatzGeneral3}
ds^2 = (1 + k|z|)^2 \left(dt^2 - dz^2\right) - (1 + k|z|)^2 \left[e^V \sum\limits_{i = 1}^{N - 3} dx_i^2 + e^{- (N - 3)V}dy^2 \right]~.
\end{equation}
For this case, instead of $z$, it is convenient to introduce the new orthogonal to the brane coordinate $r$ by the relation:
\begin{equation} \label{z-r}
1 - k|z| = e^{-a|r|}~.
\end{equation}
Note that the brane is placed at the origins of $z$ and of the new coordinate $r$ as well. In terms of $r$ the main metric {\it ansatz} (\ref{MetricAnsatzGeneral3}) of this section takes the form:
\begin{equation} \label{metricN}
ds^2 = e^{2a|r|}\left[ dt^2 - e^V \sum\limits_{i = 1}^{N - 3} dx_i^2  - e^{- (N - 3)V}dy^2 \right] - dr^2~.
\end{equation}
This metric, together with the oscillating exponents $V(t, |r|)$ that describe standing waves, contains the familiar to the standard brane models warp factor, $e^{2a|r|}$, where the constant $a$ corresponds to the brane width.

%%%%%%%%%%%%%%%%%%%%%%%%%%%%%%%%%%%%%%%%%%%%%%%%%%%%%%%%%%%%%%%%%%%%%%%%%

\subsection{5D braneworlds with ghost scalars}

In 5D ($N = 5$) the metric {\it ansatz} (\ref{metricN}) has the form:
\begin{equation} \label{metric5D}
ds^2 = e^{2a|r|}\left( dt^2 - e^V dx_1^2 - e^V dx_2^2 - e^{-2V}dy^2 \right) - dr^2~,
\end{equation}
where the curvature scale $a > 0$. For this metric the system of 5D Einstein-Klein-Gordon equations has the standing wave solution \cite{5D-ghost-1, 5D-ghost}:
\begin{equation} \label{V,phi}
V(t,|r|) \sim \phi (t, |r|) \sim \sin (\omega t) f(|r|)~.
\end{equation}
Here $\omega$ is the frequency of standing waves and
\begin{equation} \label{fsol}
f(|r|) = e^{-2a|r|} J_2\left( \frac{\omega}{|a|} e^{-a|r|}\right) ~,
\end{equation}
where $J_2$ is the second-order Bessel functions of the first kind.

Let us review the localization of different matter fields on the brane for the solution (\ref{metric5D}) with (\ref{V,phi}) and (\ref{fsol}).

%%%%%%%%%%%%%%%%%%%%%%%%%%%%%%%%%%%%%%%%%%%%%%%%%%%%%%%%%%%%%%%%%%%%%%%%%

\subsubsection{Localization of classical particles}

The 5D geodesic equation of motion for a classical particle, or a photon, has the form:
\begin{equation} \label{Geo}
\frac{d^2 x^A}{dk^2} + \Gamma^A_{BC} \frac{d x^B}{dk}\frac{d x^C}{dk} = 0~,
\end{equation}
where $k$ is the parameter of trajectory. The first integrals of this system for the brane metric (\ref{metric5D}) are \cite{5D-ghost-all}:
\begin{eqnarray} \label{sol-matter}
\frac{dx}{dk} &=& v e^{-a|r|-V/2}~, \nonumber \\
\frac {dt}{dk} &=& e^{-a|r|} ~, \\
\frac 12 \left(\frac{d r}{d k}\right)^2 &=& E - a(1 - v^2) |r|~,\nonumber
\end{eqnarray}
where the constants $v$ and $E >0$ correspond to the component of the particles velocity along the brane and to the energy per unit mass, respectively. The term $a (1 - v^2) |r|$ in the last expression of (\ref{sol-matter}) plays the role of the trapping gravitational potential.

To find a connection of the parameter $k$ with the proper time one can insert (\ref{sol-matter}) into the definition of the interval (\ref{metric5D}),
\begin{equation}
ds^2 = e^{2a|r|} dt^2 - e^{2a|r|+ V} dx^2 - dr^2 = \left[ (1 + 2a|r|) (1-v^2) - E \right]dk^2~.
\end{equation}
From this expression it is clear that on the brane ($r=0$) we have $E = 1$ for photons and $0 < E < 1$ for massive particles. The motion towards the extra dimension $r$ is possible when
\begin{equation}
E - a (1 - v^2) |r| \geq 0~,
\end{equation}
and for any energy $E$ there exists the maximal distance in the bulk,
\begin{equation}
|r|_{max} \sim \frac Ea~,
\end{equation}
the particle can reach, i.e. the classical particles are trapped on the brane.

In the standard brane approaches \cite{brane-1, brane-2, brane-3, brane-4} with the decreasing warp factor ($a < 0$) localization is achieved due to the fact that the extra space actually is finite. Here the increasing warp factor, $e^{2a|r|}$, creates the potential well that confines particles. Another point is that in (\ref{sol-matter}) the influence of the oscillating exponents of (\ref{metric5D}) is negligible, i.e. the anisotropy of the background metric for classical particles is hidden.

%%%%%%%%%%%%%%%%%%%%%%%%%%%%%%%%%%%%%%%%%%%%%%%%%%%%%%%%%%%%%%%%%%%%%%%%%

\subsubsection{Localization of scalar fields}

Consider 5D massless scalar field on the background metric (\ref{metric5D}) with the action:
\begin{equation} \label{Sphi}
S = - \frac 12 \int \sqrt{g} dx^4dr ~ g^{MN}\partial_M\Phi \partial_N\Phi~.
\end{equation}
Separating the variables,
\begin{equation}\label{Solution1}
\Phi \left( {t,x,y,z,r} \right) = \phi (x^\nu) \varsigma (r) \sim e^{i\left(Et + p_xx + p_yy + p_zz \right)} \varsigma (r)~,
\end{equation}
i.e. on the brane for the scalar zero modes we assume the standard dispersion relation,
\begin{equation} \label{Dispersion}
E^2 = p_x^2 + p_y^2 + p_z^2~,
\end{equation}
the action integral (\ref{Sphi}) can be split in two parts:
\begin{equation} \label{Sphi1}
S = - \frac 12 \int d^4x \left(\partial_\alpha \phi \partial^\alpha \phi\int dr e^{2a|r|}\varsigma^2 - \phi^2\int dr e^{4a|r|}\varsigma'^2\right)~.
\end{equation}

The time averaged (see \ref{Averages}) Klein-Gordon equation for the extra dimensional factor $\varsigma (r)$ has the asymptotic solutions \cite{5D-ghost-sca},
\begin{eqnarray} \label{solution-s}
\left. \varsigma (r)\right|_{r \to 0} &\sim& const ~, \nonumber \\
\left. \varsigma (r)\right|_{r \to \infty} &\sim& e^{-4a|r|}~.
\end{eqnarray}
From these expressions it is clear that the integrals by $r$ in (\ref{Sphi1}) are convergent, what means that the scalar field $\Phi$ is localized on the brane.

\begin{figure}[ht]
\begin{center}
\includegraphics[width=0.7\textwidth]{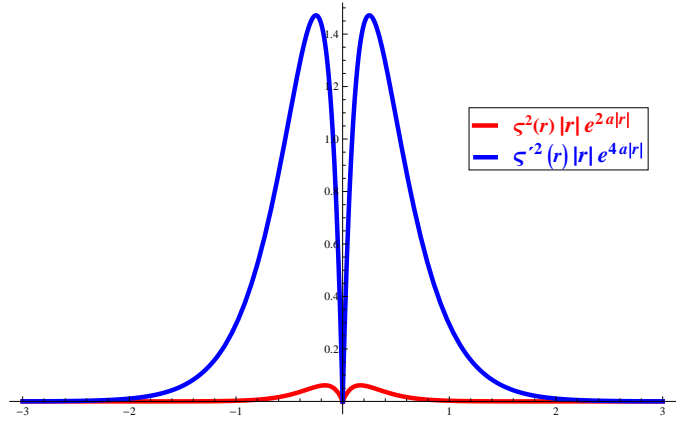}
\caption{Convergence of integrand functions in (\ref{Sphi1}).}
\label{Skalar_2}
\end{center}
\end{figure}

Localization of scalar zero modes on the brane can be shown also exactly \cite{5D-ghost-num}.
%%%%%%%%%%%%%%%%%%%%%%%%%%%%%%%%%%%%%%%
\begin{figure}[ht]
\begin{center}
\includegraphics[width=0.7\textwidth]{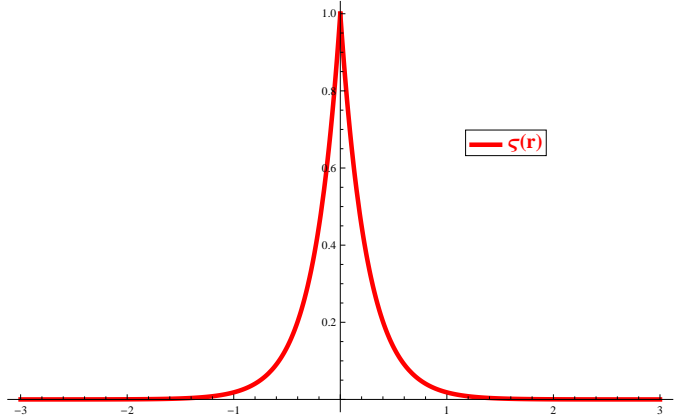}
\caption{Scalar zero mode's profile in the bulk.}
\label{Skalar_1}
\end{center}
\end{figure}
%%%%%%%%%%%%%%%%%%%%%%%%%%%%%%%%%%%%%
Fig. \ref{Skalar_1} shows the numerical solution to the time averaged Klein-Gordon equation for the boundary conditions,
\begin{equation}
\varsigma(0) = 1~, ~~~~~ \varsigma'(0) = - 4a~,
\end{equation}
and for the following values of the parameters:
\begin{equation}
\omega =3.38 \sim a~, ~~~~~ E = 0.01 \ll a~.
\end{equation}
We see that $\varsigma (r)$ rapidly falls off from the brane to zero. The integrals over $r$ in the action (\ref{Sphi1}) are convergent if the integrand functions decrease stronger than $1/r$. Fig. \ref{Skalar_2} shows that the products of the integrand functions on $r$ indeed decrease.
%%%%%%%%%%%%%%%%%%%%%%%%%%%%%%%%%%%%%%%%%%%%%%%%%
%%%%%%%%%%%%%%%%%%%%%%%%%%%%%%%%%%%%%%%%%%%%%%%%%%%%%

%%%%%%%%%%%%%%%%%%%%%%%%%%%%%%%%%%%%%%%%%%%%%%%%%%%%%%%%%%%%%%%%%%%%%%%%%

\subsubsection{Localization of gravitons}

It is known that the transverse traceless graviton modes obey the equation of massless scalar fields in a curved background. Indeed, consider the metric fluctuations:
\begin{equation} \label{metric-h}
ds^2 = e^{2a|r|}\left( g_{\mu\nu} + h_{\mu\nu}\right)dx^\mu dx^\nu - dr^2~,
\end{equation}
where $g_{\mu\nu}$ is the metric tensor of the 4D part of (\ref{metric5D}):
\begin{equation} \label{metric-4}
g_{\mu\nu} = \left( 1, - e^{V}, - e^{V}, - e^{-2V} \right) ~.
\end{equation}

Close to the brane ($V \approx 0$) for time averages of oscillatory exponents one can use the approximation:
\begin{equation}
\left\langle e^{V(t,|r|)} \right\rangle \approx 1 + \left\langle V (t,|r|)\right\rangle~.
\end{equation}
Thus the functions $\left\langle V \right\rangle$ can be regarded as $r$-dependent additive terms of $h_{\mu\nu}$. Then the equations of motion for the fluctuations $h_{\mu\nu}$,
\begin{equation}\label{graviton}
\frac{1}{\sqrt g}~\partial_M \left( \sqrt g g^{MN}\partial_N h_{\mu\nu} \right) = 0~,
\end{equation}
are equivalent to the Klein-Gordon equation of motion of a scalar field if we replace $h_{\mu\nu}$ with $\Phi$. Accordingly, the condition of localization of spin-$2$ graviton field is equivalent to that of spin-$0$ scalar field considered above \cite{5D-ghost-al, 5D-ghost-all}.

%%%%%%%%%%%%%%%%%%%%%%%%%%%%%%%%%%%%%%%%%%%%%%%%%%%%%%%%%%%%%%%%%%%%%%%%%

\subsubsection{Localization of vector fields}

The action of the 5D massless $U(1)$ gauge field has the form:
\begin{equation}\label{VectorAction}
S = - \frac14\int d^5x\sqrt g~ g^{MN}g^{PR}F_{MP}F_{NR}~,
\end{equation}
where
\begin{equation} \label{F-vec}
F_{MP} = \partial_M A_P - \partial_P A_M~.
\end{equation}
Separating the variables,
\begin{eqnarray}\label{VectorDecomposition}
A_t(x^C) &=& \upsilon (r)~a_t(x^\nu)~, \nonumber \\
A_i(x^C) &=& e^{V(t,r)} \upsilon (r)~a_i(x^\nu)~, ~~~~~ (i=x,y)\nonumber \\
A_z(x^C) &=& e^{-2V(t,r)} \upsilon (r)~a_z(x^\nu)~, \\
A_r(x^C) &=& 0~,\nonumber
\end{eqnarray}
the 5D action (\ref{VectorAction}) can be written as:
\begin{equation}\label{VectorAction1}
S = - \frac14\int d^4x \left[ F_{\alpha\beta}F^{\alpha\beta}\int dr \upsilon ^2 - 2A_\alpha A^\alpha \int dr e^{2a|r|}\upsilon'^2\right] ~.
\end{equation}

Using the standard dispersion relation for free particle on the brane (\ref{Dispersion}) the time averaged 5D Maxwell equations yields the asymptotic solutions \cite{5D-ghost-vec}:
\begin{eqnarray} \label{solution-v}
\left. \upsilon (r)\right|_{r \to 0} &\sim& const ~, \nonumber \\
\left. \upsilon(r)\right|_{r \to \infty} &\sim&  e^{-2a|r|}~.
\end{eqnarray}
For this asymptotically decreasing factor the extra dimension integrals in (\ref{VectorAction1}) are convergent, i.e. zero mode of the $U(1)$ vector field is localized on the brane.

%%%%%%%%%%%%%%%%%%%%%%%%%%%%%%%%%%%%%%%%%%%%%%%%%%
\begin{figure}[ht]
\begin{center}
\includegraphics[width=0.7\textwidth]{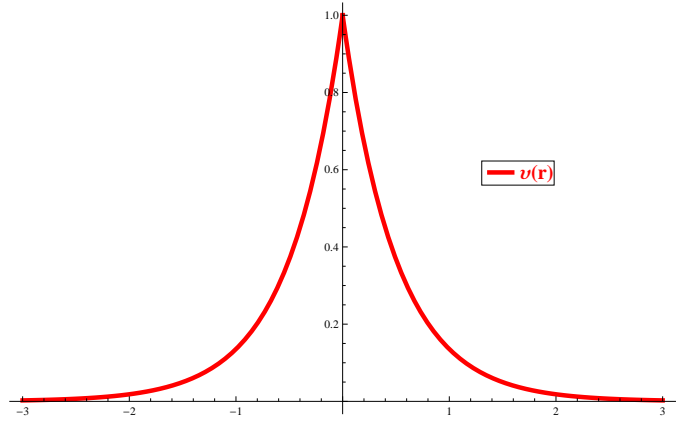}
\caption{Vector zero mode's profile in the bulk.}
\label{Vect_1}
\end{center}
\end{figure}
%%%%%%%%%%%%%%%%%%%%%%%%%%%%%%%%%%%%%%%%%%%%%%%%%

\begin{figure}[ht]
\begin{center}
\includegraphics[width=0.7\textwidth]{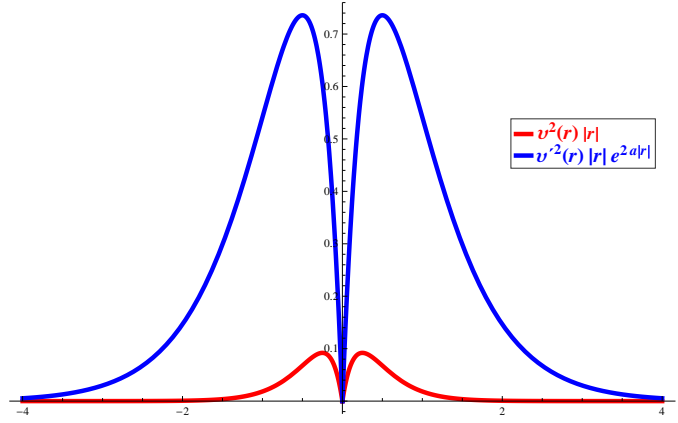}
\caption{Convergence of integrand functions in the vector field action (\ref{VectorAction1}).}
\label{Vect_2}
\end{center}
\end{figure}

This result can be verified using numerical solutions \cite{5D-ghost-num}. Fig. \ref{Vect_1} displays the shape of $\upsilon (r)$ close to the brane for the boundary conditions:
\begin{equation}
\upsilon (0) = 1~, ~~~~~ \upsilon'(0) = - 2a~.
\end{equation}
One can see that the probability of photon to leave the brane falls down to zero in the bulk. Fig. \ref{Vect_2} shows that the products of the integrand functions in (\ref{VectorAction1}) on $r$ also decrease. So the integrals over $r$ in (\ref{VectorAction1}) are convergent.
%%%%%%%%%%%%%%%%%%%%%%%%%%%%%%%%%%%%%%%%
%%%%%%%%%%%%%%%%%%%%%%%%%%%%%%%%%%%%%%%

%%%%%%%%%%%%%%%%%%%%%%%%%%%%%%%%%%%%%%%%%%%%%%%%%%%%%%%%%%%%%%%%%%%%%%%%%

\subsubsection{Localization of massless fermions}

The 5D Dirac action for free massless fermions is:
\begin{equation}\label{SpinorAction}
S = \int d^5x \sqrt g ~i\overline \Psi \left(x^A\right) \Gamma ^MD_M \Psi \left(x^A\right)~,
\end{equation}
where 5D gamma-matrix has the components:
\begin{eqnarray}\label{GammaMatrices-5D}
\Gamma^t &=& e^{ - a|r|}~ \gamma^t ~, \nonumber \\
\Gamma^i &=& e^{ - a|r| - V/2}~ \gamma^i~, ~~~~~ ~~~~~ (i=x,y) \nonumber \\
\Gamma^z &=& e^{ - a|r| + V}~ \gamma^z~, \\
\Gamma^r &=& i\gamma^5 ~. \nonumber
\end{eqnarray}

It is convenient to perform the chiral decomposition of 5D spinor wave function:
\begin{equation}\label{Psi}
\Psi \left(x^\nu,r\right) = \psi_L \left(x^\nu\right) \lambda (r) + \psi_R \left(x^\nu\right) \rho (r)~.
\end{equation}
where $\lambda(r)$ and $\rho(r)$ are extra dimension factors of the left and right brane fermions respectively. Using this decomposition the action (\ref{SpinorAction}) can be written as:
\begin{eqnarray}\label{SpinorAction1}
S &=& \int d^4x \left\{\overline \psi_L i\gamma^\mu \partial_\mu \psi_L \int dr e^{3a|r|}\lambda^2 + \overline \psi_R  i\gamma^\mu \partial_\mu \psi_R \int dr e^{3a|r|}\rho^2 +\right. \\
&+&\left. \overline \psi_R \psi_L \int dr e^{4a|r|}\rho \left[\lambda' + 2a~ sgn(r) \lambda \right] - \overline \psi_L \psi_R \int dr e^{4a|r|}\lambda \left[\rho' + 2a~ sgn(r) \rho \right]\right\}~.\nonumber
\end{eqnarray}

The asymptotic solutions to the time averaged 5D Dirac equation for the extra dimension factors $\lambda(r)$ and $\rho(r)$ are \cite{5D-ghost-al, 5D-ghost-all}:
\begin{eqnarray}\label{L-0}
&\rho (r)|_{r \to \pm 0} \sim 0 ~, ~~~~~&\lambda (r)|_{r \to \pm 0} \sim  e^{ - 2a|r|}~, \nonumber \\
&\rho (r)|_{r \to \pm \infty } \sim e^{ -2 a|r|}~, ~~~~~ &\lambda (r)|_{r \to \pm \infty } \sim e^{ - 3a|r|}~.
\end{eqnarray}
According to (\ref{L-0}) right fermion zero modes does not exist on the brane (since $\rho (0) = 0$). Also $\rho (r)$ at the infinity decreases only as $e^{ -2a|r|}$ and the integral over $r$ in the second term of (\ref{SpinorAction1}) diverges. So wave functions for right fermions actually are not normalizable.

%%%%%%%%%%%%%%%%%%%%%%%%%%%%%%%%%%%%%%%%%%%%%%%%%%%%%%
\begin{figure}[ht]
\begin{center}
\includegraphics[width=0.7\textwidth]{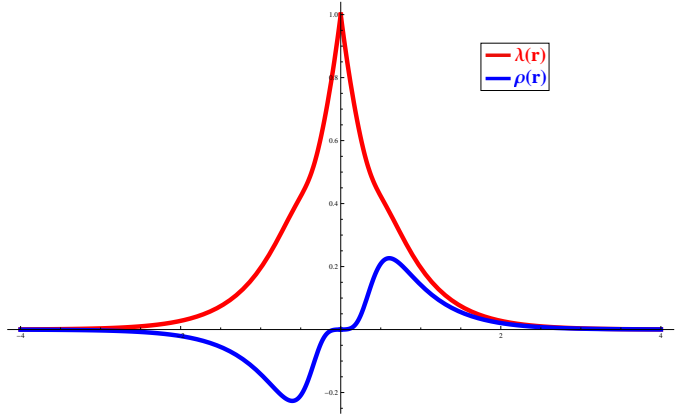}
\caption{Profiles of the left and right fermion wavefunctions in the bulk.}
\label{Ferm_1}
\end{center}
\end{figure}
%%%%%%%%%%%%%%%%%%%%%%%%%%%%%%%%%%%%%%%%%%%%%%%%%%%%%%%

\begin{figure}[ht]
\begin{center}
\includegraphics[width=0.6\textwidth]{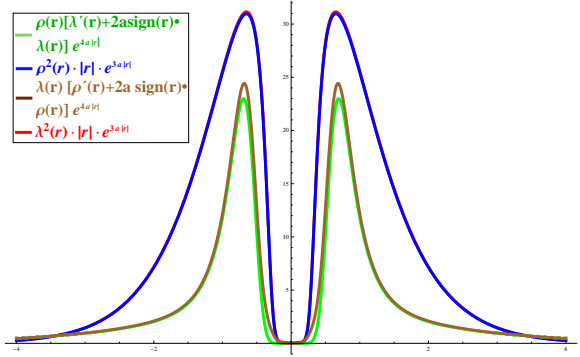}
\caption{Convergence of bulk integrals in (\ref{SpinorAction1}).}
\label{Ferm_2}
\end{center}
\end{figure}

Fig. \ref{Ferm_1} displays profiles of $\lambda(r)$ and $\rho(r)$ for the boundary conditions \cite{5D-ghost-num}:
\begin{equation}
\rho (0) = 0, ~~~~~\lambda (0) = 1~.
\end{equation}
Integrals over $r$ in the spinor field action (\ref{SpinorAction1}) will be convergent if integrand functions decrease stronger than $1/r$. This feature for all terms of (\ref{SpinorAction1}) is demonstrated on the Fig. \ref{Ferm_2}. From these figures we see that $\lambda(r)$ has maximum on the brane and decreases in the bulk. While $\rho (r)$ has maximum in the bulk outside the brane. So in this model left massless fermions are localized on the brane and right fermions are localized in the bulk.

%%%%%%%%%%%%%%%%%%%%%%%%%%%%%%%%%%%%%%%%%%%%%%%%%%
%%%%%%%%%%%%%%%%%%%%%%%%%%%%%%%%%%%%%%%%%%%%%%%%%%

%%%%%%%%%%%%%%%%%%%%%%%%%%%%%%%%%%%%%%%%%%%%%%%%%%%%%%%%%%%%%%%%%%%%%%%%%

\subsubsection{Hierarchy of fermion masses}

\begin{figure}[ht]
\begin{center}
\includegraphics[width=0.7\textwidth]{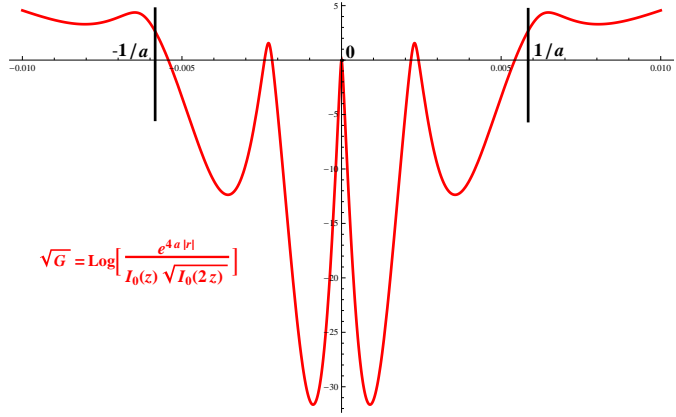}
\caption{Logarithmic profile of the determinant of (\ref{metric5D}) for $\omega /a\approx 10$.}
\label{det}
\end{center}
\end{figure}

The metric (\ref{metric5D}) describes the brane located at a node of the standing wave, which can be considered as the 4D space-time 'island', where the matter particles are assumed to be bound. Then the replication of fermions families might be connected with the localization of fermionic modes around different 'islands' \cite{5D-ghost-mass}. If one fine tune the parameters, $\omega /a\approx 10.02$, the time averaged matric (\ref{metric5D}) will exhibit three nodes of the bulk standing wave distributed symmetrically with respect the central node. Fig. \ref{det} shows the shape of the determinant of (\ref{metric5D}) along the extra dimension $r$ \cite{5D-ghost-mass}.
%%%%%%%%%%%%%%%%%%%%%%%%%%%%%%%%%%%%%%%%%%%%%%%%%%%
%%%%%%%%%%%%%%%%%%%%%%%%%%%%%%%%%%%%%%%%%%%%%%%%%%%%
Correspondingly, there appear several fermionic modes which are 'stuck' at different points in the brane of the width $\sim 1/a$ \cite{Split-1, Split-2}.

Consider the action of 5D massive fermions,
\begin{equation} \label{S}
S = \int d^5 x \sqrt{G} \left[ \frac{i}{2} \overline{\Psi} \Gamma^{M} D_M \Psi - \frac{i}{2} D_M  \overline{\Psi } \Gamma^{M} \Psi - M \overline{\Psi} \Psi \right]~,
\end{equation}
The 5D fermion wavefunction can be decomposed into chiral left and right ones (\ref{Psi}), where the extra dimension factors $\lambda(r)$ and $\rho(r)$ of the 4D left and right fermion wavefunctions are even and odd functions of the extra coordinate, respectively. Then (\ref{S}) takes the form:
\begin{eqnarray}\label{action2}
S_\psi &=& i \int dr \sqrt{^5g}\frac{\lambda^2}{2} \int d^4 x \left(\overline{\psi}_L \Gamma^\mu D_\mu \psi_L - D_\mu \overline{\psi}_L \Gamma^{\mu} \psi_L \right) + \nonumber \\
&+& i \int dr \sqrt{^5g}\frac{\rho^2}{2} \int d^4 x \left(\overline{\psi}_R \Gamma^\mu D_\mu \psi_R - D_\mu \overline{\psi}_R \Gamma^{\mu} \psi_R \right) + \nonumber \\
&+& i \int dr \sqrt{^5g}\frac{\lambda \rho}{2} \int d^4 x \left(\overline{\psi}_L \Gamma^\mu D_\mu \psi_R + \overline{\psi}_R \Gamma^\mu D_\mu \psi_L - \right.\\
&&~~~~~~~~~~~~~~~~~ - \left.D_\mu \overline{\psi}_R \Gamma^{\mu} \psi_L- D_\mu \overline{\psi}_L \Gamma^{\mu} \psi_R \right) - \nonumber \\
&-& \int dr \sqrt{^5g}\left( M \rho \lambda -\frac{\rho \lambda'}{2} + \frac{\lambda \rho'}{2} \right) \int d^4 x \left(\overline{\psi}_L \psi_R+\overline{\psi}_R \psi_L \right)~. \nonumber
\end{eqnarray}

The 5D Dirac equation for the 4D fermions of the mass $m$ with zero momentum along the brane gives the solutions \cite{5D-ghost-mass}:
\begin{eqnarray} \label{lambda,rho}
\lambda(r) &=& \Bigg{[} \left(C_u \frac{M}{\mu} - C_d \frac{m}{\mu} \right) \sinh(\mu r) + C_u \cosh(\mu r) \Bigg{]}e^{-2ar} ~, \nonumber \\
\rho(r) &=& \Bigg{[} \left(C_u \frac{m}{\mu} - C_d \frac{M}{\mu} \right) \sinh(\mu r) + C_d \cosh(\mu r) \Bigg{]}e^{-2ar} ~,
\end{eqnarray}
where $C_u$ and $C_d$ are the integration constants and
\begin{equation}
\mu \equiv \sqrt{M^2 - m^2} ~.
\end{equation}
The solutions (\ref{lambda,rho}) lead to the localization of 4D fermions on the brane \cite{5D-ghost-mass}. Indeed, the first two terms in (\ref{action2}) are convergent over $r$ and the third term vanishes, because $\lambda (r)$ is an even and $\rho (r)$ is an odd function of $r$.

\begin{figure}[ht]
\begin{center}
\includegraphics[width=0.7\textwidth]{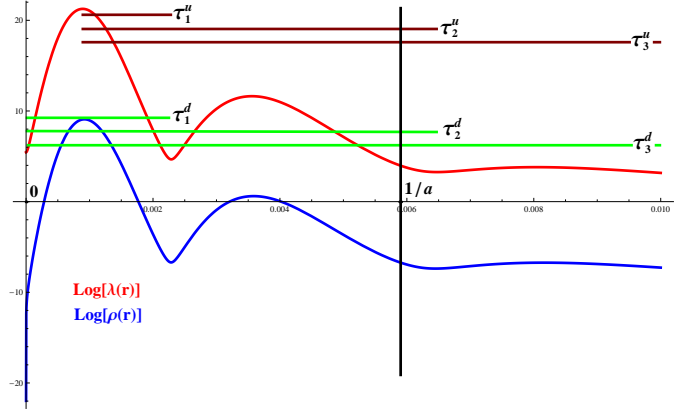}
\caption{Logarithmic profiles of the left and right fermions along the extra dimension.}
\label{modes}
\end{center}
\end{figure}

Consider the last term in (\ref{action2}), which corresponds to the 4D fermion masses \cite{5D-ghost-mass}:
\begin{eqnarray} \label{m}
m^{fer} &=& \int dr \left( M \rho\lambda + \frac{\lambda \rho'}{2}-\frac{\rho \lambda'}{2} \right) \frac{e^{4a|r|}}{I_0(f) \sqrt{I_0(2f)}}= \nonumber \\
&=& \left[\left(C_u^2 + C_d^2\right)m - 2 C_uC_dM\right]\int_0^{\infty} \frac{dr}{I_0(f) \sqrt{I_0(2f)}} ~,
\end{eqnarray}
where $I_0$ is modified Bessel function of zero order and the function $f(r)$ is done in (\ref{fsol}). The fermion families can be connected with the existence of several peaks of wave functions which are located at different points in the bulk (see Fig. \ref{modes}).

%%%%%%%%%%%%%%%%%%%%%%%%%%%%%%%%%%%%%%%%%%%%%%%%
%%%%%%%%%%%%%%%%%%%%%%%%%%%%%%%%%%%%%%%%%%%%%%%%

The values of the distances between the peaks and nodes of the left and right fermion wavefunctions can be used to explain observed mass spectrum of the fermion families. For example, for three down quarks ($C_u=0$ in (\ref{m})),
\begin{equation}\label{mi}
m_i^d = m_d ~ C_d^2 \int_0^{\tau_i^d} \frac{dr}{I_0(f) \sqrt{I_0(2f)}}~,
\end{equation}
the upper limits of integration in (\ref{mi}) acquire the values:
\begin{equation}
\tau_1^d = 0.0021~, ~~~ \tau_2^d = 0.0065~, ~~~\tau_3^d = 0.0750~,
\end{equation}
and we obtain the observed mass ratios.

In the case of up quarks one can assume $C_d=0$ in (\ref{m}) and write
\begin{equation}\label{mi2}
m_i^u = m_u ~ C_u^2 \int_{0.0009}^{\tau_i^u} \frac{dr}{I_0(f) \sqrt{I_0(2f)}}~,
\end{equation}
where the limits of integration (brown lines on Fig. \ref{modes}) are:
\begin{equation}
\tau_1^u = 0.0021~, ~~~ \tau_2^u = 0.0065~, ~~~\tau_3^u = 0.2250~.
\end{equation}
Then for (\ref{mi2}) we are able to obtain the observed mass ratios of up quarks also.

To obtain the absolute values for mass spectrum we first need to specify the physical units. The 5D fermionic mass $M$ should exceed the 4D mass of any quark. Hence, in the minimal model we can take,
\begin{equation}
M \simeq a \simeq m_t \simeq 172~GeV~.
\end{equation}
Then the observed mass spectrum of all down and up quarks can be reproduced if
\begin{equation}
C_d \simeq 860~GeV^{1/2}~, ~~~~~ C_u \simeq 181~GeV^{1/2}
\end{equation}
in (\ref{mi}) and (\ref{mi2}).

%%%%%%%%%%%%%%%%%%%%%%%%%%%%%%%%%%%%%%%%%%%%%%%%%%%%%%%%%%%%%%%%%%%%%%%%%

\subsection{6D braneworlds with ghost scalars}

For six dimensional models ($N = 6$) the metric {\it ansatz} (\ref{metricN}) can be written as \cite{6D-ghost-mid}:
\begin{equation}\label{BS-OutsideMetricAnsatz}
ds^2 = e^{2ar}\left[ dt^2 - e^V\left( dx^2 + dy^2 + dz^2 \right) \right] - dr^2 - R_0^2e^{2ar - 3V}d\theta ^2~,
\end{equation}
where $a \neq 0$ and $R_0 > 0$ are the constants, and the function $V (t,r)$ depends only on time $t$ and on the extra radial polar coordinate $r \geq 0$. When $V=0$ this metric describes a string-like topological defect at $r=0$ in the 6D space-time \cite{Od}. The {\it ansatz} (\ref{BS-OutsideMetricAnsatz}) differs from the metric considered in the paper \cite{6D-ghost-SSA} in the sense that it symmetrically considers the brane coordinates $x$, $y$ and $z$.

When the warp factor $a$ in (\ref{BS-OutsideMetricAnsatz}) relates to the 6D cosmological constant as $\Lambda_{6} = -10a^2$ the system of 6D Einstein-Klein-Gordon equations has the standing wave solution (\ref{V,phi}) with
\begin{equation}\label{BS-Function-B(r)-withZn}
f(r) \sim e^{ - 5ar/2} J_{5/2}\left( Z_n e^{ - ar} \right)~,
\end{equation}
where the constant $Z_n = \omega/a$ denotes $n$-th zero of the Bessel function $J_{5/2}$. In this model the 2D extra space consists with $(n-1)$ concentric circles (all of them sharing the same center coinciding with the origin of the extra 2D space), where the function $J_{5/2}\left( Z_n e^{ - ar} \right)$ vanishes. These circles are the nodes of the standing wave in the 2D extra space and can be considered as the circular islands around the string-like defect where matter particles can be bound.

Below we shall review the localization mechanism of different matter fields on the string-like defect when $Z_1 \approx 5.76$. In this case bulk circular standing waves have only two nodes, at $r = 0$ and at $r = \infty$.

%%%%%%%%%%%%%%%%%%%%%%%%%%%%%%%%%%%%%%%%%%%%%%%%%%%%%%%%%%%%%%%%%%%%%%%%%

\subsubsection{Localization of scalar fields}

Consider 6D scalar field with the action (\ref{Action-Phi}). Separating the variables
\begin{equation}\label{SF-Solution}
\Phi (t,x,y,z,r,\theta) = e^{i\left(Et - p_nx^n \right)}\sum\limits_{l,m} \phi_m (r)e^{il\theta }~,
\end{equation}
for the standard dispersion relation (\ref{Dispersion}) the time averaged Klein-Gordon equation on the background metric (\ref{BS-OutsideMetricAnsatz}) has the following asymptotic solutions for the $S$-wave zero modes ($m=l=0$) \cite{6D-ghost-mid}:
\begin{eqnarray}
\phi_0 (r) |_{r \to 0} &\to& 5a + e^{ -5ar}~, \nonumber \\
\phi_0 (r) |_{r \to \infty} &\to& e^{ -5ar} ~.
\end{eqnarray}
It is obvious that for these solutions extra dimension integrals in (\ref{Action-Phi}) are convergent, i.e. scalar field zero modes are localized on the brane.

%%%%%%%%%%%%%%%%%%%%%%%%%%%%%%%%%%%%%%%%%%%%%%%%%%%%%%%%%%%%%%%%%%%%%%%%%

\subsubsection{Localization of vector fields}

The 6D $U(1)$ gauge field action has the form:
\begin{equation}\label{VectorAction 6D}
S = - \frac{1}{4}\int d^6x\sqrt {-g}~ g^{MN}g^{PR}F_{MP}F_{NR}~,
\end{equation}
where $F_{MP}$ is defined in (\ref{F-vec}). Consider a solution to the corresponding 6D Maxwell equations in the form:
\begin{eqnarray}\label{VectorFieldDecomposition}
A_t (x^C) &=& a_t(x^\nu)\sum \limits_{l,m} \rho _m (r) e^{il\theta} ~, \nonumber \\
A_k (x^C) &=& e^{V(t,r)} a_k (x^\nu) \sum \limits_{l,m} \rho_m (r)e^{il\theta}~, ~~~~~(k=x,y,z)\nonumber \\
A_r (x^C) &=& a_r(x^\nu)\sum \limits_{l,m} \rho_m (r) e^{il\theta }~, \\
A_\theta (x^C) &=& 0 ~. \nonumber
\end{eqnarray}
For the $S$-waves and the dispersion relation (\ref{Dispersion}), the asymptotic solutions for the zero modes ($l=m=0$) are \cite{6D-ghost-mid}:
\begin{eqnarray}\label{VF-Solution-rho-asymptotics}
\left.\rho _0 (r)\right|_{r \to 0} &\to& 3a + e^{ -3ar} ~, \nonumber \\
\left. \rho _0 (r)\right|_{r \to \infty} &\to& e^{ -3ar}~.
\end{eqnarray}
This result shows that on the background metric (\ref{BS-OutsideMetricAnsatz}) the extra dimension integrals in (\ref{VectorAction 6D}) are convergent, i.e. vector fields are localized on the brane.

%%%%%%%%%%%%%%%%%%%%%%%%%%%%%%%%%%%%%%%%%%%%%%%%%%%%%%%%%%%%%%%%%%%%%%%%%

\subsubsection{Localization of fermions}

The action for 6D massless fermions is,
\begin{equation}\label{FF-DiracAction 6D}
S = \int d^6x \sqrt { -g} i\bar \Psi \Gamma ^AD_A \Psi~ ,
\end{equation}
where in the space-time (\ref{BS-OutsideMetricAnsatz}) nonzero components of gamma matrices are:
\begin{eqnarray}\label{FF-Relation-Gamma-gamma}
  \Gamma^t &=& e^{ - ar}\gamma^t ~, \nonumber \\
  \Gamma^k &=& e^{ - ar - V/2}\gamma^k ~, ~~~~~(k =x,y,z) \nonumber \\
  \Gamma^r &=& \gamma^r ~, \\
  \Gamma^\theta  &=& \frac {1}{R_0} e^{ - ar + 3V/2} \gamma^\theta ~. \nonumber
\end{eqnarray}
After separation of the variables,
\begin{equation}\label{FF-Spinor-Psi-decomposition 6D}
\Psi (x^A) = \psi (x^\nu )\sum\limits_{l,m} \alpha _m(r)e^{il\theta }~,
\end{equation}
one can find the $S$-wave zero mode solutions to the 6D Dirac equation:
\begin{equation} \label{FF-ZeroMode}
\left. \alpha_0(r) \right|_{r \to 0} \sim \left. \alpha _0(r) \right|_{r \to \infty} \to e^{-5ar/2}~.
\end{equation}
This means that extra dimension integrals in (\ref{FF-DiracAction 6D}) are convergent and fermions are also localized on the string-like defect (\ref{BS-OutsideMetricAnsatz}) \cite{6D-ghost-mid}.

%%%%%%%%%%%%%%%%%%%%%%%%%%%%%%%%%%%%%%%%%%%%%%%%%%%%%%%%%%%%%%%%%%%%%%%%

\section{\bf Solutions with $k \neq 0$}
\label{k ne 0}

In the case when the constants $k$ in (\ref{MetricAnsatzGeneral}) is non-zero the system of equations (\ref{BulkScalarFieldEquation}) and (\ref{SystemOfEquationsForMetricFunction}) have solutions for the following values of the exponents:
\begin{equation}\label{a,b}
c = - \frac{N - 3}{N - 2}~, ~~~~~ b = \frac{2}{N - 2}~.
\end{equation}
In this case the metric (\ref{MetricAnsatzGeneral}) obtanes the form:
\begin{eqnarray}\label{MetricGeneral}
ds^2 &=& \frac {1}{(1 + k|z|)^{(N-3)/(N-2)}} e^S \left(dt^2 - dz^2\right) - \nonumber \\
&-&(1 + k|z|)^{2/(N-2)} \left[e^V \sum\limits_{i = 1}^{N - 3} dx_i^2  + e^{- (N - 3)V}dy^2 \right]~.
\end{eqnarray}
When $k < 0$ this metric, together with the singularity at $|z|=0$ (where the brane is placed), has the horizons at $|z| = -1/k$. At these points some components of Ricci tensor get infinite values, while all gravitational invariants, for example the Ricci scalar,
\begin{equation}\label{RicciScalar}
  R = 2\left( S' + \frac{N - 1}{N - 2}k \right)e^{- S}\delta (z) + \left(1 + k|z| \right)^{(N - 3)/(N - 2)}e^{- S}\left( S'' - \ddot S \right)~,
\end{equation}
are finite. This resembles the situation with the Schwarzschild black hole, however, the determinant of (\ref{MetricGeneral}) is zero at $|z| = -1/k$. As the result, nothing can cross these horizons and for the brane observer the extra space $z$ is effectively finite.

%%%%%%%%%%%%%%%%%%%%%%%%%%%%%%%%%%%%%%%%%%%%%%%%%%%%%%%%%%%%%%%%%%%%%%%%%

\subsection{Domain wall in $N$ dimensions}

First of all we want to mention the simplest case of (\ref{MetricGeneral}) without the scalar field and metric functions:
\begin{equation}
\phi = V = S = 0 ~,
\end{equation}
corresponding to the static $N$-dimensional domain wall \cite{GMT}:
\begin{equation}
ds^2 = (1 + k|z|)^{- \frac{N - 3}{N - 2}}\left( dt^2 - dz^2 \right) - (1 + k|z|)^{\frac{2}{N - 2}}\left( \sum\limits_{i = 1}^{N - 3} dx_i^2 + dy^2 \right)~.
\end{equation}
In this case the brane tensions in (\ref{SystemOfEquationsForBraneEnergyMomentumTensor}) are:
\begin{eqnarray}
\tau_{t} &=& - 2k~, \nonumber \\
\tau_{x_1} &=& ... = \tau _{x_{(N - 3)}} = \tau_{y} = - \frac{N - 3}{N - 2}~k~, \\
\tau_{z} &=& 0~. \nonumber
\end{eqnarray}

%%%%%%%%%%%%%%%%%%%%%%%%%%%%%%%%%%%%%%%%%%%%%%%%%%%%%%%%%%%%%%%%%%%%%%%%%

\subsection{4D waves bounded by a domain wall}

In four dimensions ($N=4$), taking the positive $k$ to avoid the horizon singularities, the metric {\it ansatz} (\ref{MetricGeneral}) takes the form \cite{GMS}:
\begin{equation}\label{Metric-4}
ds^2 = \frac {e^S}{\sqrt{1 + k|z|}} \left(dt^2 - dz^2\right) - (1 + k|z|) \left(e^V dx^2  + e^{-V}dy^2 \right)~.
\end{equation}
This matric appears to be some combination of the domain wall solution \cite{Taub, Vilenkin, Ip-Si} and the colliding plane wave solutions \cite{Yurt, Fe-Ib, Griff} and describes a plane symmetric, standing gravi-scalar waves bounded by a domain wall.

In the case of normal bulk scalar field, when $\epsilon = +1$ in (\ref{action-N}), for the background metric (\ref{Metric-4}) the system of Einstein and Klein-Gordon equations gives the solutions:
\begin{eqnarray} \label{solution-4}
V(t,z) &=& C_1J_0 \left(\frac{\omega}{k} + \omega z \right) \cos (\omega t)~, \nonumber \\
\phi (t,z) &=& \frac{C_1}{2} J_0 \left( \frac{\omega}{k}+\omega z \right) \sin (\omega t) ~, \\
S(z) &=& C_2 +\frac{C_1 \omega ^2}{4k^2} (1 + kz)^2 \left[ J_0^2\left( \frac{\omega}{k}+\omega z \right) + \right. \nonumber \\
&+& \left. 2 J_1^2 \left( \frac{\omega}{k}+\omega z \right) - J_0 \left( \frac{\omega}{k}+\omega z \right)J_2 \left( \frac{\omega}{k}+\omega z \right) \right] ~,\nonumber
\end{eqnarray}
where $C_1$ and $C_2$ are the integration constants and $J_0$, $J_1$ and $J_2$ are ordinary Bessel functions of zeroth, first and second order, respectively.

From (\ref{solution-4}) one finds that the metric function, $V(t,z)$, and the scalar field, $\phi (t,z)$, had the same spatial dependence but their time oscillations are $\pi /2$ out of phase. One could view this as the energy of the oscillation passing back and forth between the scalar and gravitational fields. Since the field energy of the scalar field can be localized this suggests that for this solution one might be able to define a local gravitational energy. However, since this solution is not asymptotically flat and the domain wall is not a localized source, one cannot define global gravitational field energy via surface integrals over effective energy-momentum tensors \cite{Steph, BPR, Eh-Ku, MTW}.

There is a lot in common between the solution (\ref{solution-4}) and the simple electromagnetic standing wave between two infinite conducting planes. The exponentially increasing Newtonian potential,
\begin{equation} \label{Phi}
\Phi (z) = \frac 12 \left[g_{00}(z) - g_{00}(z=0)\right] = \frac{\exp[S(z)]}{2\sqrt{1 + kz}} -\frac{1}{2} \approx \frac{\exp[C_1 \omega z / \pi ]}{\sqrt{kz}}~,
\end{equation}
which traps the oscillatory parts of the gravitational field, may be thought of as the second (soft) plane in conjunction with the (hard) plane of the domain wall at $z=0$.

%%%%%%%%%%%%%%%%%%%%%%%%%%%%%%%%%%%%%%%%%%%%%%%%%%%%%%%%%%%%%%%%%%%%%%%%%

\subsection{Gravi-ghost waves bounded by the brane}

Now we consider the case with the phantom scalar field ($\epsilon = -1$) when the metric function $S$ in (\ref{MetricGeneral}) is zero. In this case the $N$-dimensional equations (\ref{BulkScalarFieldEquation}) and (\ref{SystemOfEquationsForMetricFunction}) have the following solutions \cite{GMT}:
\begin{eqnarray}
V &=& C \sin (\omega t)J_0(X)~, \nonumber \\
\phi &=& \frac C2 \sqrt {M^{N-2}(N - 2)(N - 3)} \sin (\omega t)J_0(X)~,
\end{eqnarray}
where $C$ is the integration constant and the argument of $J_0$ is defined as:
\begin{equation}\label{X}
X = \frac{|\omega|}{|k|}(1 + k|z|)~.
\end{equation}
Imposing the boundary condition,
\begin{equation}
V |_{|z| = 0} = 0 ~,
\end{equation}
the equations (\ref{SystemOfEquationsForBraneEnergyMomentumTensor}) for the brane tensions will give the solution:
\begin{eqnarray}
\tau_t &=& - 2k, \nonumber \\
\tau_{x_1} &=& \tau_{x_2} = ... = \tau_{x_{N - 3}} = -\frac{N - 3}{N - 2} k + V'~, \nonumber \\
\tau_y &=& -\frac{N - 3}{N - 2} k - (N - 3)V'~, \\
\tau_z &=& 0~. \nonumber
\end{eqnarray}

%%%%%%%%%%%%%%%%%%%%%%%%%%%%%%%%%%%%%%%%%%%%%%%%%%%%%%%%%%%%%%%%%%%%%%%%%

\subsection{Standing wave braneworlds with normal source}

The general solution to the $N$-dimensional equations (\ref{BulkScalarFieldEquation}), (\ref{SystemOfEquationsForMetricFunction}) and (\ref{SystemOfEquationsForBraneEnergyMomentumTensor}) for the case of the normal bulk scalar field ($\epsilon = +1$) is done by \cite{GMT}:
\begin{eqnarray} \label{V,phi,S}
V &=& \left[ C_1 \sin (\omega t) + C_2\cos (\omega t) \right]\left[ C_3J_0(X) + C_4Y_0(X)\right]~, \nonumber \\
\phi &=& \frac 12 \sqrt {M^{N-2}(N - 2)(N - 3)} \left[ C_1\cos (\omega t) - C_2\sin (\omega t)\right]\left[ C_3J_0(X) + C_4Y_0(X) \right]~, \nonumber \\
S &=& \frac 12 (N - 2)(N - 3)X^2\left\{ C_3^2\left[ J_0(X)^2 + J_1(X)^2 - \frac 1X J_0(X)J_1(X) \right] + \right. \\
&+& C_4^2 \left[ Y_0(X)^2 + Y_1(X)^2 - \frac 1X Y_0(X)Y_1(X) \right] +  \nonumber\\
&+& \left. C_3C_4 \left[ 2\left[ J_0(X)Y_0(X) + J_1(X )Y_1(X) \right] - \frac 1X \left[ J_0(X)Y_1(X) + J_1(X)Y_0(X ) \right] \right] \right\} + C_5~, \nonumber
\end{eqnarray}
where $C_i$ ($i=1, 2, 3, 4, 5$) and $\omega$ are some constants, $J_0$, $J_1$ and $Y_0$, $Y_1$ are Bessel functions of the first and the second kind, respectively, and $X$ is defined in (\ref{X}).

Imposing the boundary conditions,
\begin{equation}
S|_{|z| = 0} = V|_{|z| = 0} = 0 ~,
\end{equation}
the system of equations (\ref{SystemOfEquationsForBraneEnergyMomentumTensor}) for the brane tensions will have the following solution:
\begin{eqnarray} \label{tau}
\tau_t &=& - 2k~, \nonumber \\
\tau_{x_1} &=& \tau _{x_2} = ... = \tau _{x_{(N - 3)}} = -\frac{N - 3}{N - 2}k - S' + V'~, \nonumber \\
\tau_y &=& -\frac{N - 3}{N - 2}k - S' - (N - 3)V'~, \\
\tau_z &=& 0~.\nonumber
\end{eqnarray}

For the 5D case the matric (\ref{MetricGeneral}) (for the negative $k$) reduces to \cite{5D-real-scal}
\begin{equation} \label{MetricAnsatz}
ds^2 = \frac {e^S}{(1 - k|r|)^{2/3}}\left( dt^2 - dr^2 \right) - (1 - k|r|)^{2/3}\left( e^V dx^2 + e^V dy^2 + e^{-2V}dz^2 \right)~.
\end{equation}
and the solutions (\ref{V,phi,S}) and (\ref{tau}) have the form:
\begin{eqnarray} \label{SysSolution}
V(t,|r|)&=& C \sin (\omega t) J_0 (X)~,\nonumber\\
\varphi(t,|r|) &=& C \cos (\omega t) J_0(X)~, \nonumber \\
S(|r|) &=& \frac 32 C^2 \left[X^2\left( J_0^2(X) + J_1^2(X) - \frac 1X J_0 (X) J_1 (X) \right) - \frac{\omega^2}{k^2} J_1^2\left(\frac{\omega}{k}\right) \right]~, \nonumber\\
\tau _t^t &=& 2k~,\\
\tau _x^x &=& \tau _y^y = \frac 23 k + \frac{3\omega^2C^2}{2k} J_1^2\left(\frac{\omega}{k}\right) + C \omega \sin(\omega t) J_1\left(\frac{\omega }{k}\right) ~,\nonumber\\
\tau _z^z &=&  \frac 23 k + \frac{3\omega^2C^2}{2k} J_1^2\left(\frac{\omega}{k}\right) - 2 C \omega \sin(\omega t) J_1\left(\frac{\omega }{k}\right)~.\nonumber
\end{eqnarray}
where $X$ is done in (\ref{X}) and $\omega /k = Z_n$ ($Z_n$ are zeros of $J_0$).

The solutions (\ref{SysSolution}) have two limiting cases corresponding to the small and the large amplitudes of the bulk standing waves:
\begin{itemize}
\item{In the first limiting case, $C \ll 1$, the amplitude and consequently the energy of the oscillations are small, i.e. the functions $V$, $S$ and $\varphi$ does not play significant role and one can consider the metric {\it ansatz} (\ref{MetricAnsatz}) without oscillatory metric functions:
\begin{equation} \label{MetricAnsatz2}
ds^2 = \frac {1}{(1 - k|r|)^{2/3}}\left( dt^2 - dr^2 \right) - (1 - k|r|)^{2/3}\left( dx^2 + dy^2 + dz^2 \right)~.
\end{equation}
This metric is 5D generalizations of the 4D domain wall solution \cite{Taub, Vilenkin, Ip-Si}. Due to the presence of the absolute value of the extra coordinate, $|r|$, the Ricci tensor at $r=0$ has $\delta$-like singularity, which corresponds to the brane tension. The metric (\ref{MetricAnsatz2}) has also new features, it exhibits the horizons at $|r| = 1/k$ and matter fields are confined inside of the fat 3-brane of the width $\sim 1/k$.}
\item{In the case of large extra space, $C \gg 1$, or when $|V/S|\ll 1$, trapping of matter fields on the brane is caused by the pressure of the bulk oscillations and not by the existence of the horizon in the extra space.}
\end{itemize}

%%%%%%%%%%%%%%%%%%%%%%%%%%%%%%%%%%%%%%%%%%%%%%%%%%%%%%%%%%%%%%%%%%%%%%%%%

\subsubsection{Localization of scalar fields}

Consider the real massless scalar field with the action (\ref{Sphi}) in the background metric (\ref{MetricAnsatz}). Separating the variables,
\begin{equation}\label{Solution-scalar}
\Phi (t,x,y,z,r) = e^{ - i\left( Et - p_xx - p_yy - p_zz \right)} \rho (r)~,
\end{equation}
the Klein-Gordon equation for the extra dimension factor $\rho(\left|r\right|)$ that obey the boundary conditions,
\begin{equation}\label{ConditionForRho(r)}
\rho'|_{|r| = 0} = 0~, ~~~~~ \rho |_{|r| \to 1/k} = 0~,
\end{equation}
have the asymptotic solutions \cite{5D-real-scal}:
\begin{eqnarray}\label{rho}
\rho_0 (r)|_{|r| \to 0} &\sim& 1 - \frac{C^2\omega^2 E^2 }{4k} J_1^2\left(\frac{\omega}{k}\right) |r|^3 ~, \nonumber \\
\rho_0 (r)|_{|r| \to 1/k} &\sim& (1 - k|r|)^{2/3} ~,
\end{eqnarray}
where $C$ is the integration constant and $\omega$ denotes the frequency of standing waves. For this zero mode solution the integral over the extra coordinate $r$ in the action (\ref{Sphi}) is finite, i.e. the scalar field is localized on the brane.

%%%%%%%%%%%%%%%%%%%%%%%%%%%%%%%%%%%%%%%%%%%%%%%%%%%%%%%%%%%%%%%%%%%%%%%%%

\subsubsection{Localization of gauge fields}

Close to the brane the wavefunction of 5D massless $U(1)$ vector field with the action (\ref{VectorAction}) can be factorized as:
\begin{eqnarray}\label{VectorFieldAnsatz}
A_t\left(x^C\right) &=& (1 - k|r|)^{-2/3}e^{S(r)} \xi (|r|) \varepsilon_t e^{i\left( Et + p_xx + p_yy + p_zz\right)}~, \nonumber \\
A_i\left(x^C\right) &=& (1 - k|r|)^{2/3}e^{V(t,r)} \rho(|r|) \varepsilon_i e^{i\left( Et + p_xx + p_yy + p_zz\right)}~, \nonumber ~~~~~~(i = x, y) \\
A_z\left(x^C\right) &=& (1 - k|r|)^{2/3}e^{-2V(t,r)} \rho(|r|) \varepsilon_z e^{i\left( Et + p_xx + p_yy + p_zz\right)}~, \\
A_r\left(x^C\right) &=& 0~, \nonumber
\end{eqnarray}
where $\varepsilon_t$, $\varepsilon_i$ and $\varepsilon_z$ are the components of the polarization 4-vector of photons on the brane. The 5D Maxwell equations for the extra dimensional factors, $\xi (|r|)$ and $\rho (|r|)$, which obey the boundary conditions:
\begin{eqnarray}\label{BoundaryConditionsOnBrane}
\left. \frac{\xi'}{\xi } \right|_{|r| \to 0} \gg S'|_{|r| \to 0}~, ~~~~~\left. \frac{\xi'}{\xi} \right|_{|r| \to 0} \gg k~, \nonumber \\
\left. \frac{\rho'}{\rho } \right|_{|r| \to 0} \gg V'|_{|r| \to 0}~, ~~~~~ \left. \frac{\rho'}{\rho} \right|_{|r| \to 0} \gg k~,
\end{eqnarray}
have the asymptotic solutions \cite{5D-real-vec}:
\begin{eqnarray}\label{SolutionAtHorizon}
\xi|_{|r| \to 0} &\sim& \rho|_{|r| \to 0}  \to C_1 - |r|, \nonumber \\
\xi|_{|r| \to 1/k} &\to& C_2 (1 - k|r|)^{2/3}~, \\
\rho|_{|r| \to 1/k} &\to& C_3 ~,\nonumber
\end{eqnarray}
where $C_1$, $C_2$ and $C_3$ are some constants. For the sharply decreasing extra dimension factors in (\ref{SolutionAtHorizon}) the 5D vector field action (\ref{VectorAction}) is integrable over the extra coordinate $r$. This means that the vector field zero modes are localized on the brane.

%%%%%%%%%%%%%%%%%%%%%%%%%%%%%%%%%%%%%%%%%%%%%%%%%%%%%%%%%%%%%%%%%%%%%%%%%

\subsubsection{Localization of fermions}

Now consider 5D spinor field zero modes with the action (\ref{SpinorAction}). In the background metric (\ref{MetricAnsatz}) the curved space-time gamma matrices are related to Minkowskian ones by the expressions:
\begin{eqnarray}\label{GammaMatrices-normal}
\Gamma^t &=& (1 - k|r|)^{1/3}e^{ -S/2}\gamma^t~,\nonumber \\
\Gamma^i &=& (1 - k|r|)^{-1/3} e^{ -V/2}\gamma^i~, ~~~~~ ~~~~~ (i=x,y) \nonumber \\
\Gamma^z &=& (1 - k|r|)^{-1/3} e^V \gamma^z~, \\
\Gamma^r &=& (1 - k|r|)^{1/3}e^{ -S/2}\gamma ^r~. \nonumber
\end{eqnarray}

Close to any $n$-th node of standing waves the spinor wavefunction can be factorized,
\begin{equation}\label{PsiNearKthNode}
\left. \Psi \left( x^A\right) \right|_{r \to r_n} \approx \psi _n \left(x^\nu\right)\rho _n(r)~,
\end{equation}
where $\rho_n (r)$ is the extra dimension scalar factor of the fermion wave function near the $n$-th node.

Consider the wavefunction (\ref{PsiNearKthNode}) for two limiting regions:
\begin{itemize}
\item{On the brane,
\begin{equation}\label{PsiOnBrane}
\left. \Psi \left( x^A\right) \right|_{r \to 0} \approx \psi _0\left( x^\nu \right)\rho _0(r)~,
\end{equation}
were $\psi _0 \left(x^\nu\right)$ corresponds to the zero mode Dirac spinor, the 5D Dirac equation has the solution:
\begin{equation}\label{RhoOnTneBrane}
\rho_0 (r) \sim e^{3C^2\omega ^2 J_1^2(\omega/a )|r|/8k}~,
\end{equation}
where $C$ is the integration constant.}
\item{Close to the horizons,
\begin{equation}\label{PsiAtHorizon}
\left. \Psi \left( x^A\right) \right|_{|r| \to 1/k} \approx \psi _h\left( x^\nu \right)\rho _h(r)~,
\end{equation}
we assuming that
\begin{equation}
\psi_h\left(x^{\nu}\right)= const~,
\end{equation}
and the 5D Dirac equation for the extra dimension factor $\rho_h (r)$ has the solution
\begin{equation}\label{RhoOnTneHorizon}
\rho_h (r) \sim \frac {1}{(1 - k|r|)^{1/3}}~.
\end{equation}
}
\end{itemize}
The extra dimension space of the model is effectively finite and for the wavefunction with the asymptotes (\ref{RhoOnTneBrane}) and (\ref{RhoOnTneHorizon}) the integrals over $r$ in (\ref{SpinorAction}) are convergent, i.e. the zero mode fermion is localized on the brane \cite{5D-real-ferm}.

%%%%%%%%%%%%%%%%%%%%%%%%%%%%%%%%%%%%%%%%%%%%%%%%%%%%%%%%%%%%%%%%%%%%%%%%%

\section{\bf The 6D model with normal source}
\label{6D normal}

The example of the braneworld with $B \ne 0$ in (\ref{MetricAnsatzGeneral}) is the model \cite{6D-SSA}:
\begin{equation} \label{metric-6}
ds^2 = e^{2ar}\left( dt^2 - e^V dx^2 - e^V dy^2 - e^{-3V}dz^2 \right) - dr^2 - R_0^2e^{a_1 r + V}d\theta^2 ~,
\end{equation}
where $a$ and $a_1$ are real constants ($a_1 \ne 2a$) and the radial coordinate $r$ is defined in (\ref{z-r}). The range of the variables $r$ and $\theta$ in (\ref{metric-6}) are $0 \leq r < \infty$ and $0 \leq \theta < 2\pi$, respectively. This metric {\it ansatz} is a combination of metrics describing 6D global string-like defect \cite{Oda2, Gregory, Kol-Kar} and 6D standing wave braneworld \cite{6D-ghost-mid} with anisotropic warping of the three brane spatial coordinates through the terms $e^{V(t,r)}$ and $e^{ -3 V(t,r)}$.

The oscillatory metric function $V$ in (\ref{metric-6}) is done by
\begin{equation} \label{V(t,r)}
V(t,r) = \sin(\omega t) f(r)~,
\end{equation}
where
\begin{equation} \label{bessel1sol_r}
f(r) = C_1 e^{- dr/2} J_{-d/2 a} \left(\frac{\omega}{a} e^{- ar}\right) + C_2 e^{-dr/2} J_{d/2a} \left(\frac{\omega}{a} e^{- ar}\right)~.
\end{equation}
Here $C_{1}$ and $C_{2}$ are integration constants and $J_{\pm d/2a}$ are the first kind Bessel functions of the orders $\pm d/2a$ with
\begin{equation}
d = \frac{11}{3}a + \frac 23 a_1~.
\end{equation}
The function (\ref{bessel1sol_r}) reduces to (\ref{BS-Function-B(r)-withZn})\cite{6D-ghost-mid} if $d = 5a$ and $C_1 = 0$. Note that both functions $J_{\pm d/2a}$ in (\ref{bessel1sol_r}) are regular at the origin and at infinity. Depending on the relation between $\omega$, $a$ and $d$ the functions $J_{\pm d/2a}$ converge for both $a>0$ or $a<0$, enabling solutions with the decreasing and increasing warp factors.

The requirement that the function (\ref{V(t,r)}) is zero on the brane quantizes waves frequency and $\omega$ may be expressed by the $n$-th zero ($Z_n = \omega /a$) of $J_{- d/2a}$ or $J_{d/2a}$ depending if we take $C_1$ or $C_2$ equal to zero in $(\ref{bessel1sol_r})$.

It can be shown that for some values of parameters of the model all components of the energy-momentum tensor of the bulk scalar field are positive and main energy conditions also are satisfied, i.e. it is possible to obtain the solution when bulk scalar field is not ghost-like. This can be demonstrated for the cases where the constants $d$ and $a$ have the same and opposite signs.

%%%%%%%%%%%%%%%%%%%%%%%%%%%%%%%%%%%%%%%%%%%%%%%%%%%%%%%%%%%%%%%%%%%%%%%

\subsection{Same sign for $d$ and $a$}

\begin{figure}[ht]
\begin{center}
\includegraphics[width=0.7\textwidth]{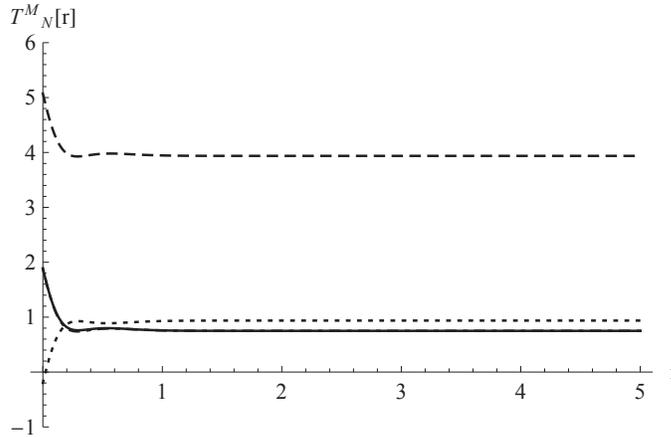}
\caption[Figure 1:]{Profile of $\langle T_{N}^{M} \rangle$ when $d$ and $a$ have the same sign.}
\label{fig:1}
\end{center}
\end{figure}

The choice
\begin{equation}
d = 4a
\end{equation}
will imply $a_1 = d/2$ and the solution (\ref{bessel1sol_r}) will depends on $J_2$ only. Fig. \ref{fig:1} shows the time averaged components of the energy-momentum tensor for this case when
\begin{equation}
C_2 = a = 1~, ~~~~~ \omega = 5.13 ~.
\end{equation}
The dot-dashed line represents $\langle T^x_x\rangle = \langle T^y_y\rangle = \langle T^z_z\rangle$, the doted one represents $\langle T^r_r\rangle$, the dashed line represents $\langle T^\theta_\theta\rangle$ and finally, the filled line represents the energy density $\langle T^t_t \rangle$. As one can see all these quantities (except of the part of $T_r^r$) are positive, but it is not possible to say that this is a normal matter once the dominant energy condition is violated. However, it is not an exotic source once the null, strong and weak energy conditions are satisfied.
%%%%%%%%%%%%%%%%%%%%%%%%%%%%%%%%%%%%%%%%%%%%%%%%%%%%%%%
%%%%%%%%%%%%%%%%%%%%%%%%%%%%%%%%%%%%%%%%%%%%%%%%%%%%%%%%%
To have a normal matter source it is necessary to consider an anisotropic cosmological constant \cite{Arc-Hub},
\begin{center}\label{Lambda}
$\Lambda = \left(
\begin{array}{c l r}
\Lambda \eta_{\mu \nu} & &\\
& \Lambda_{5} & \\
& & \Lambda_{6}
\end{array} \right)$~,
\end{center}
where $\eta_{\mu \nu}$ is the metric of the brane.

\begin{figure}[ht]
\begin{center}
\includegraphics[width=0.7\textwidth]{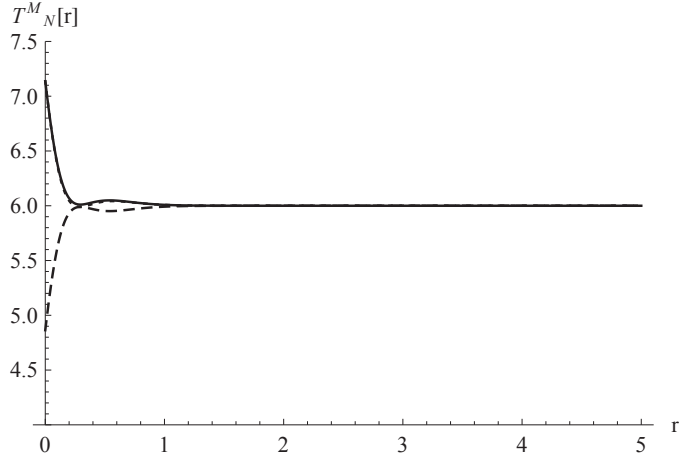}
 \caption[Figure 1:]{Profile of $\langle T_{N} ^{M} \rangle$ for anisotropic cosmological constant.}
\label{fig:2}
\end{center}
\end{figure}

Fig. \ref{fig:2} displays components of the energy-momentum tensor when
\begin{equation}
\Lambda = - \frac 14 \left( a_1^2 + 6a a_1\right)~, ~~~~~ \Lambda_5 = - 2 a a_1 ~, ~~~~~ \Lambda_6 = - 4 a^2 ~.
\end{equation}
The dotted line represents the spatial components of the energy-momentum tensor, except the $r$ component, which is represented by the shaded line and the filled line represents the temporal component. As one can see all these quantities are positive and all the energy conditions are satisfied.

%%%%%%%%%%%%%%%%%%%%%%%%%%%%%%%%%%%%%%%%%%%%%%%%%
%%%%%%%%%%%%%%%%%%%%%%%%%%%%%%%%%%%%%%%%%%%%%%%%%

%%%%%%%%%%%%%%%%%%%%%%%%%%%%%%%%%%%%%%%%%%%%%%%%%%%%%%%%%%%%%%%%%%%%%%%%%

\subsection{Opposite signs for $d$ and $a$}

\begin{figure}[ht]
\begin{center}
\includegraphics[width=0.7\textwidth]{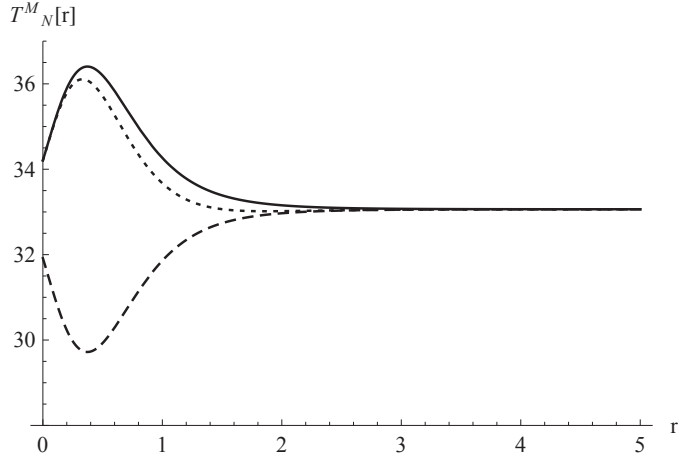}
\caption[Figure 1:]{Profile of $\langle T_{N} ^{M} \rangle$ when $d$ and $a$ have opposite signs.}
\label{fig:3}
\end{center}
\end{figure}

Consider
\begin{equation}
d = -4a ~,
\end{equation}
which will give $a_1 = - 23a/2$. Again, if the cosmological constant is isotropic it is possible to find solution with all energy-momentum tensor components positive, but it would not possible to obey the dominant energy condition. So we need a solution with the anisotropic cosmological constant (\ref{Lambda}):
\begin{equation}
\Lambda = - 6a^2 + \frac 32 a a_1 ~, ~~~~~ \Lambda_5 = - 6a^2 + 2 a a_1 - \frac 14 a_1^2 ~, ~~~~~ \Lambda_6 = - 10 a^2 + \frac 14 a_1^2~.
\end{equation}
Fig. \ref{fig:3} shows the time averaged components of the energy-momentum tensor for $C_{2} = 0$. As above the filled line represents the energy density, the dotted one gives $\langle T^{x} _{x} \rangle = \langle T^{y} _{y} \rangle = \langle T^z_z \rangle = \langle T^\theta_\theta \rangle$ and the dashed line represents the $\langle T^r_r \rangle$ component. As one can see all these quantities are positive and the energy-momentum tensor assures the dominant energy condition. Therefore, again we obtained a standing wave solution generated by normal matter.

%%%%%%%%%%%%%%%%%%%%%%%%%%%%%%%%%%%%%%%%%%%%%%%
%%%%%%%%%%%%%%%%%%%%%%%%%%%%%%%%%%%%%%%%%%%%%%

%%%%%%%%%%%%%%%%%%%%%%%%%%%%%%%%%%%%%%%%%%%%%%%%%%%%%%%%%%%%%%%%%%%%%%%%%

\subsection{Localization problem}

\begin{figure}[ht]
\begin{minipage}[b]{0.48 \linewidth}
   \fbox{\includegraphics[width=\linewidth]{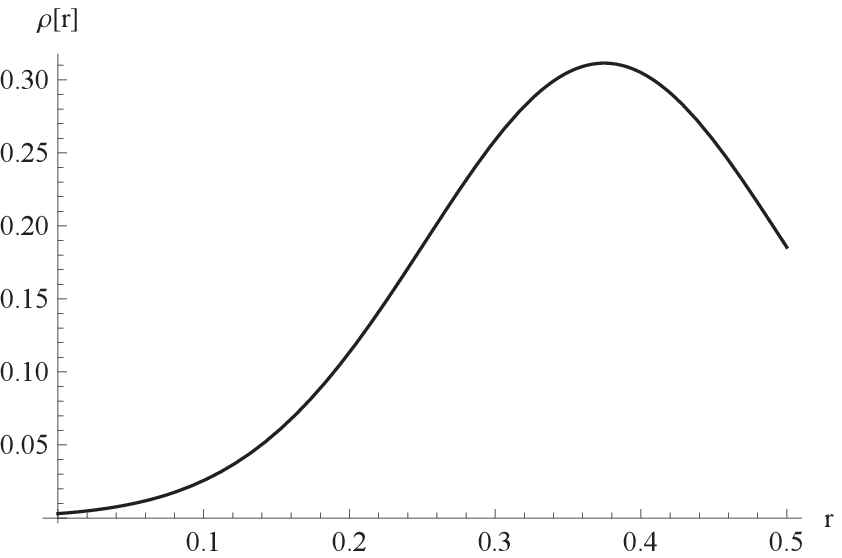}}\\
   \caption{Profile of $\rho (r)$ for $a = 1$.}
   \label{fig:4}
\end{minipage}\hfill
\begin{minipage}[b]{0.49 \linewidth}
   \fbox{\includegraphics[width=\linewidth]{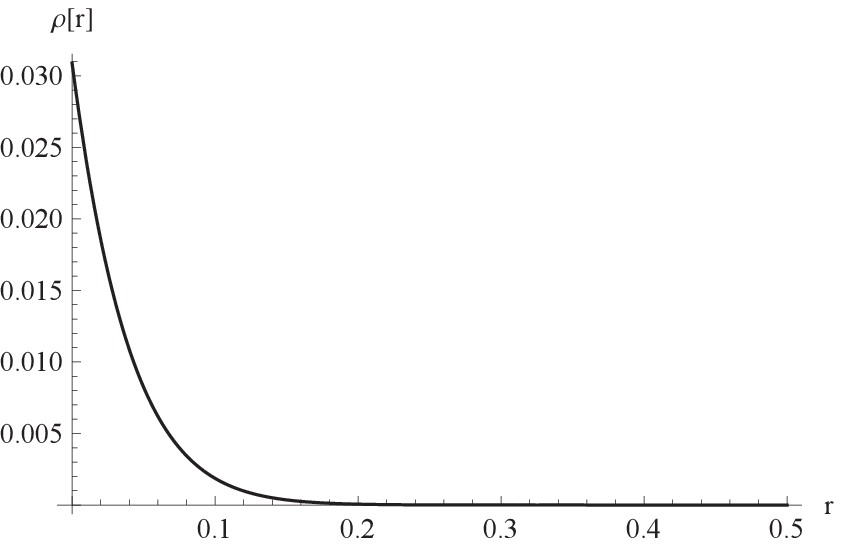}}\\
   \caption{Profile of $\rho (r)$ when $a = -1$.}
   \label{fig:5}
\end{minipage}
\end{figure}

Consider the localization of a 6D scalar field with the action (\ref{Action-Phi}) on the background metric (\ref{metric-6}). For simplicity it can be taken $C_{1} = 0$ and $d = 4a$ in (\ref{bessel1sol_r}). Once we do this, the $V(r,t)$ will depends on the first kind Bessel function $J_2$. Then we assume $\omega/|a| = 5.13$, which corresponds to the first zero of $J_{2}$. Separating the variables,
\begin{equation}
\label{scalar_var_sep}
\Phi (t,x,y,z,r,\theta) = e^{i\left(Et - p_xx - p_yy -p_zz \right)}\sum\limits_{l,m} \rho_m (r)e^{il\theta }~,
\end{equation}
and using the dispersion relation (\ref{Dispersion}), the time averaged Klein-Gordon equations have the solutions for the asymptotic values of zero mode extra dimension factor:
\begin{eqnarray}
\rho (r)|_{r \to 0} &\to& e^{-(2a + a_1/4)r}~, \nonumber \\
\rho (r)|_{r \to \infty} &\to& e^{- (33a/8 + a_1/4) r}~,
\end{eqnarray}
which is convergent for either $a > 0$ or $a < 0$. For the case $a=1$, as can be seen in Fig. \ref{fig:4}, the extra part of the scalar zero-mode wave function $\rho$ has a minimum at $r=0$, increases and then fall off. For $a=-1$ the function has a maximum at $r = 0$ and it rapidly falls off as we move away from the brane, see Fig. \ref{fig:5}. So extra dimension integrals in (\ref{Action-Phi}) are convergent, i.e. scalar field zero modes are localized on the brane for increasing or decreasing warp factor.

%%%%%%%%%%%%%%%%%%%%%%%%%%%%%%%%%%%%%%%%%%%%%%%%%%
%%%%%%%%%%%%%%%%%%%%%%%%%%%%%%%%%%%%%%%%%%%%%%%%%%

At the end of this paragraph note that localization of vector bosons and fermions on the background (\ref{metric-6}) is very similar to the model (\ref{BS-OutsideMetricAnsatz}) \cite{6D-ghost-mid}, except that there the Bessel function considered is $J_{5/2}$, and here it is used $J_2$.

%%%%%%%%%%%%%%%%%%%%%%%%%%%%%%%%%%%%%%%%%%%%%%%%%%%%%%%%%%%%%%%%%%%%%%%%%

\section{\bf Cosmological solutions}
\label{Cosmology}

In this section we review cosmological solutions within a generalized standing wave braneworlds modeled when the main metric {\it ansatz} (\ref{MetricAnsatzGeneral}) contains the extra time-depended scale factor $a(t)$.

Usually in standing wave braneworlds with ghost scalar field it is assumed the oscillating metric function $V(t,|r|)$ to be proportional to the bulk scalar field $\phi(t,|r|)$. In general, it is possible to relax this restrictive assumption, since the system of Einstein and Klein-Gordon equations is self-consistent also for the case when the time derivatives of these functions are proportional, $\dot V(t,|r|) \sim \dot \phi(t,|r|)$. So one can consider more general solution with the extra proportional to time term, which is useful in cosmological applications.

%%%%%%%%%%%%%%%%%%%%%%%%%%%%%%%%%%%%%%%%%%%%%%%%%%%%%%%%%%%%%%%%%%%%%%%%%

\subsection{Brane isotropization mechanism}

Considering the 5D standing wave braneworld model with the metric \cite{Cosmo-iso}:
\begin{equation} \label{metricA}
ds^2 = e^{2A(r)}\left[dt^2 - a^2(t)\left(e^V dx^2 + e^V dy^2 + e^{-2V}dz^2\right) - dr^2\right]~.
\end{equation}
This {\it ansatz} generalizes the metric (\ref{metric5D}) to the case when $A(r) \ne 2a|r|$ and also contains extra time-dependent scale factor $a(t)$, which multiplies the warped spatial coordinates allowing them to evolve in time (expanding or contracting). Apart from the bulk gravitational waves of the original model this allows us to study cosmological effects, in particular, a possible isotropization mechanism that led the Universe to shed its anisotropy \cite{bwisotropization}.

In general, to study a model with anisotropic backgrounds one must have a complete solution to the bulk and brane field equations. This is not an easy task \cite{bwcosmolanisot} and there are very few anisotropic non-stationary braneworlds that present a complete solution \cite{bwisotropization, bwcosmolanisot, Iso-1, Iso-2, Iso-3, Iso-4, Iso-5}. For the model (\ref{metricA}) the Einstein equation has the exact solution of the form \cite{Cosmo-iso}:
\begin{eqnarray} \label{a,A}
a(t) &=& e^{Ht}~, \nonumber \\
A(r) &=& \ln \left(\frac Hb \mbox{\rm sech}\left[ H(r-r_0) \right]\right)~.
\end{eqnarray}
Here $H$ is a constant and the second parameter $b$ relates to the 5D cosmological constant, $\Lambda = 6b^2$. Then separating the variables,
\begin{equation} \label{separation}
\phi (t,r) \sim V(t,r) = \epsilon(t) \chi(r) e^{- 3A/2}~,
\end{equation}
the Klein-Gordon equation for the bulk scalar field $\phi (t,r)$ reduces to a couple of ordinary differential equations:
\begin{eqnarray}\label{chi}
\chi'' - \left[U (r) - \Omega^2 \right] \chi = 0~, \nonumber \\
\label{e} \ddot \epsilon + 3H\dot \epsilon + \Omega^2 \epsilon = 0~,
\end{eqnarray}
where $U(r)$ is the analog quantum mechanics P\"oschl-Teller potential,
\begin{eqnarray}\label{U}
U (r) = \frac 94 H^2 - \frac {15}{4} H^2 {\rm sech}^2\left[ H (r-r_0) \right]~,
\end{eqnarray}
and $\Omega$ and $r_0$ are some constants.

The differential equation for $\chi$ in (\ref{chi}) turns out to describe a known eigenvalue problem with a mixed spectrum. Namely, there are a continuum of KK states starting at $\Omega= 3H/2$ and two bound states: a ground state with $\Omega=0$ and another one with $\Omega = \sqrt{2}H$, separated by a gap that is determined by the asymptotic value:
\begin{equation}
U (\infty) = \frac 94 H^2~.
\end{equation}
The equation for $\chi$ in (\ref{chi}) possesses the general solution:
\begin{equation} \label{chisol}
\chi (r) = C_1 P^\mu_{3/2} \left(\tanh\left[H(r-r_0)\right]\right) + C_2 Q^\mu_{3/2} \left(\tanh\left[H(r-r_0)\right]\right)~,
\end{equation}
where $C_1$ and $C_2$ are integration constants and $P^{\mu}_{3/2}$ and $Q^{\mu}_{3/2}$ are the Legendre functions of first and second kind of the degree $\nu=3/2$ and order
\begin{equation} \label{mu}
\mu = \sqrt{\frac {\Omega^2}{H^2} - \frac 94} ~.
\end{equation}

The first discrete state, the zero mode
\begin{equation}\label{O0}
\chi_0 (r) = c_0 {\rm sech}^{3/2}\left[ H (r-r_0) \right]~,
\end{equation}
where $c_0$ is a normalization constant, corresponds to the ground state with $\Omega=0$, order $\mu=3/2$, and the energy
\begin{equation}
E_0 = -\frac 94 H^2 ~.
\end{equation}

The second bound state corresponds to an excited mode with $\Omega = \sqrt{2}H$, order $\mu = 1/2$, and the energy
\begin{equation}
E_1 = -\frac 14 H^2~,
\end{equation}
and has the following form:
\begin{equation}
\chi_1 (r) = c_1 \sinh\left[H(r-r_0)\right]{\rm sech}^{3/2}\left[ H(r-r_0) \right]~,
\end{equation}
where $c_1$ also is a normalization constant.

Finally, for the continuum of KK massive modes the order (\ref{mu}) becomes purely imaginary,
\begin{equation}
\mu = i\sqrt{\frac {\Omega^2}{H^2} - \frac 94} ~.
\end{equation}

Now consider the equation for time dependent part of the bulk scalar field, $\epsilon(t)$, in (\ref{chi}). This equation describes a damped oscillator which has three different solutions depending on the relation between the constants $H$ and $\Omega$:
\begin{itemize}
\item{In under-damping case, $\Omega^2 > 9H^2/4$, the solution of (\ref{chi}) for $\epsilon(t)$ is:
\begin{equation}\label{undere}
\epsilon(t) \sim e^{-\frac 32 Ht} \sin\left(\omega t + \delta\right)~. ~~~~~ \left(\omega = \sqrt{\Omega^2 - \frac 94 H^2}\right)
\end{equation}
From this solution it follows that oscillations exponentially decay with time, which leads to an isotropic 5D metric for a 3-brane with de Sitter symmetry.}
\item{The solution for the critical damping, $\Omega^2 = 9H^2/4$, reads:
\begin{equation}\label{crite}
\epsilon(t) = e^{-\frac 32 Ht}\left(c_1 t + c_2 \right)~,
\end{equation}
where $c_1$ and $c_2$ are the integration constants. There is the same effect of isotropization of 3-brane as in the previous case.}
\item{The over-dumped case, $\Omega^2 < 9H^2/4$, possesses the following solution:
\begin{equation}\label{overe}
\epsilon(t) = e^{-\frac 32 Ht}\left(c_1 e^{\varpi t} + c_2 e^{-\varpi t}\right)~, ~~~~~\left(\varpi = \sqrt{\frac 94 H^2-\Omega^2}\right)
\end{equation}
where $c_1$ and $c_2$ are arbitrary constants determined by initial conditions.}
\end{itemize}

We see that, in general, the solution for the time evolution of the metric function $V(t,r)$ expressed by (\ref{separation}) exponentially yields an isotropic 5D metric of the form:
\begin{equation} \label{isotropicmetric}
ds^2 = e^{2A(r)}\left[dt^2 - dr^2 - a^2(t)\left(dx^2 + dy^2 + dz^2 \right) \right]~,
\end{equation}
where the functions $A(r)$ and $a(t)$ are done in (\ref{a,A}). So the anisotropic metric (\ref{metricA}) will exponentially evolve to an isotropic 5D metric (\ref{isotropicmetric}) since all the solutions for $\epsilon(t)$ exponentially vanish in time for any values of the integration constants. It is worth noticing that together with the metric function $V$, the scalar field $\phi$ also exponentially disappears as a consequence of (\ref{separation}), rendering a completely geometric de Sitter thick brane. A physical interpretation of this dissipation can be that the anisotropic energy of the 3-brane rapidly leaks into the bulk through the nontrivial components of the projected to the brane non-local Weyl tensor. The bulk becomes less isotropic, at the same time the anisotropic braneworld exponentially isotropizes by itself and the phantom scalar field vanishes.

%%%%%%%%%%%%%%%%%%%%%%%%%%%%%%%%%%%%%%%%%%%%%%%%%%%%%%%%%%%%%

\subsection{Dimensional reduction}

It is known that some braneworld models can provide us with a geometrical mechanism of dimensional reduction supported by a curved extra dimension \cite{KEK}. In this section we consider the dynamical dimensional reduction in generalized standing waves braneworlds with ghost scalar fields \cite{Cosmo-red}.

Consider the 5D standing waves braneworld (\ref{metric5D}) in which the novel scale factors $a_1(t)$ and $a_2(t)$ are introduced \cite{Cosmo-red}:
\begin{equation} \label{Metric-F}
ds^2 = e^{2a|r|}\left[ dt^2 - a_1^2(t) e^{V(t,r)} \left(dx^2 + dy^2 \right) - a_2^2(t)e^{-2V(t,r)}dz^2\right] - dr^2~.
\end{equation}
For this metric {\it ansatz} the 5D Einstein equation with the cosmological constant $\Lambda = 6 a^2$ have the exact solutions:
\begin{eqnarray}\label{expsoln}
V(t,r) &\sim& \sin (\omega t) e^{-2a|r|} J_2\left( \frac{\omega}{|a|} e^{-a|r|} \right)~, \nonumber \\
a_1(t) &\sim& e^{Ht}~,\\
a_2(t) &\sim& e^{-2Ht}~,\nonumber
\end{eqnarray}
where $H$ is a constant. For these solutions the Klein-Gordon equation in the space-time (\ref{Metric-F}) gives:
\begin{equation} \label{solutions}
\phi(t,r) \sim V(t,r) + 2Ht ~.
\end{equation}

From (\ref{expsoln}) it is obvious that when the constant $H$ is positive the space-time (\ref{Metric-F}) expands exponentially in the $x$ and $y$ directions and squeezes in the $z$ direction. This means that in a macroscopic time interval the brane surface at $r=0$ will shrink into a 2-brane, i.e. the 3-brane will effectively have two space-like dimensions. At the same time the amplitude of the ghost scalar field in (\ref{solutions}) will increase with time.

In the case of negative $H$, in the space (\ref{Metric-F}) the $z$-distances will expand and the $(x-y)$-plane will shrink, leading to a 1-string. In this case we shall have just one spatial dimension in the 3-brane. Simultaneously the amplitude of the ghost field in (\ref{solutions}) will decrease in time.

So starting with the anisotropic 5D metric (\ref{Metric-F}) and leaving it evolve for large times, certain spatial dimensions of the 3-brane will shrink to zero-size while others will expand in an accelerated way. This mechanism of dynamical asymmetric dimensional reduction of multi-dimensional surfaces could be useful for string models when obtaining a 4D isotropic expanding space-time from a higher-dimensional anisotropic universe. An example is the generalized 6D standing wave braneworld \cite{6D-ghost-oto} with the metric \cite{Cosmo-Oto}:
\begin{equation} \label{ansatz}
ds^2 = e^S dt^2 - a(t)^2e^V \left(dx^2 + dy^2 + dz^2\right) - dr^2 - \frac{1}{a(t)^6}e^{-3V} d\theta^2~.
\end{equation}
This metric {\it ansatz} is modification of (\ref{ansatz-6D}) by the novel scale factor $a(t)$, in addition to $S(|r|)$ and $V(t,|r|)$. Solutions to the system of Einstein and Klein-Gordon equations in this case are:
\begin{eqnarray}\label{S_sol}
S (r) &=& \ln \left(1 + \frac {|r|}{a} \right)^2~, \nonumber \\
V(t,|r|) &=& \sin (\omega t) f(|r|)~, \\
a(t) &\sim& e^{H t}~, \nonumber
\end{eqnarray}
where $a$ is a constant and
\begin{equation} \label{f-cosm}
f(|r|) \sim \sin \left( a\omega \ln\left[1 + \frac{|r|}{a}\right]\right) ~.
\end{equation}
This expressions differs from the analogous solutions (\ref{Sol-6D}) \cite{6D-ghost-oto} by the new exponential scale factor $a(t)$.

For the solutions (\ref{S_sol}) the metric (\ref{ansatz}) takes the form:
\begin{eqnarray} \label{s_ansatz}
ds^2 &=& \left( 1 + \frac{|r|}{a}\right)^2 dt^2 - e^{sin(\omega t) f(|r|) + 2Ht} \left(dx^2 + dy^2 + dz^2\right) - \nonumber \\
&-& dr^2 - e^{-3 sin(\omega t) f(|r|) - 6Ht}d\theta^2~.
\end{eqnarray}
So, as for the 5D case, amplitudes of the oscillatory exponents in (\ref{s_ansatz}) increases/decreases with time depending on the sign of the constant $H$. For the positive $H$ the space-time (\ref{ansatz}) expands exponentially in the $x$, $y$ and $z$ directions, while the angle $\theta$ squeezes. This means that in a macroscopic time interval the space will effectively have three space-like dimensions,
\begin{equation} \label{m_ansatz}
ds^2 = dt^2 - e^{2H t} \left( dx^2 + dy^2 + dz^2 \right)~,
\end{equation}
i.e. spatial volume performs inflationary expansion.

Without changing of the main features of the model the number of compact extra dimensions $\theta$ in (\ref{ansatz}) can be increased \cite{GMT}. So this mechanism of dynamical dimensional reduction of multi-dimensional surfaces could be useful for wide class of string models.

%%%%%%%%%%%%%%%%%%%%%%%%%%%%%%%%%%%%%%%%%%%%%%%%%%%%%%%%%%%%%%%%%%%%%%%

\section*{Acknowledgments:}

This research was supported by the Shota Rustaveli National Science Foundation grant ${\rm ST}09\_798\_4-100$ and Javakhishvili Tbilisi State University grant.

%%%%%%%%%%%%%%%%%%%%%%%%%%%%%%%%%%%%%%%%%%%%%%%%%%%%%%%%%%%%%%%%%%%%
%%%%%%%%%%%%%%%%%%%%%%%     Appendix     %%%%%%%%%%%%%%%%%%%%%%%%%%%
%%%%%%%%%%%%%%%%%%%%%%%%%%%%%%%%%%%%%%%%%%%%%%%%%%%%%%%%%%%%%%%%%%%%

\appendix
\gdef\thesection{Appendix \Alph{section}}
\renewcommand{\theequation}{\Alph{section}.\arabic{equation}}

%%%%%%%%%%%%%%%%%%%%%%%%%%%%%%%%%%%%%%%%%%%%%%%%%%%%%%%%%%%%%%%%%%%

\section{Time averages of oscillatory functions}
\label{Averages}
\setcounter{equation}{0}

The standing waves solutions of the Einstein equations are done by the oscillating metric function,
\begin{equation} \label{V}
V(t,|z|) \sim \sin (\omega t) f(|z|)~,
\end{equation}
where $\omega$ is the frequency of the waves and $f(|z|)$ is some function depended on the extended extra dimension $z$. The function (\ref{V}) enters the equations of matter fields via exponentials:
\begin{equation} \label{e-br}
e^{bV} = \sum \limits_{n = 0}^{+\infty } \frac{\left( bV \right)^n}{n!}~,
\end{equation}
where $b$ is some constant. If $\omega$ is much larger than frequencies associated with energies of particles on the brane one can perform time averaging of oscillating exponents in the equations of matter fields. From the mathematical expression:
\begin{equation}
\frac{\omega }{2\pi} \int\limits_0^{2\pi/\omega} \left[ \sin (\omega t)\right]^m dt = \left \{
\begin{array} {lr}
0 & ( m = 2n + 1)\\
\frac {2^{-m}m!}{ \left[ (m/2)! \right]^2} & (m = 2n)
\end{array}
\right.
\end{equation}
it follows the simple formula for the time averages of (\ref{e-br}) \cite{Gr-Ry}:
\begin{equation} \label{e-u}
\left\langle e^{bV} \right\rangle = \sum \limits_{n = 0}^{+\infty } \frac{ f(|z|)^{2n}}{2^{2n}\left( n! \right)^2} = I_0(f(|z|)) ~,
\end{equation}
where $I_0$ is the modified Bessel function of the zero order. To simplify equations of various matter fields on the brane within a standing wave braneworld model it is useful also the following equalities for time averages of various oscillatory functions:
\begin{equation}\label{AdditionalFacts}
\left\langle V \right\rangle = \left\langle V' \right\rangle = \left\langle \frac{\partial V}{\partial t} \right\rangle  = \left\langle \frac{\partial V}{\partial t}e^{ - V} \right\rangle = 0~,
\end{equation}
where prime denotes the derivative with respect to the extra coordinate $z$.

%%%%%%%%%%%%%%%%%%%%%%%%%%%%%%%%%%%%%%%%%%%%%%%%%%%%%%%%%%%%%%%%%%%%%%%%%

\end{document}